\documentclass[fleqn,usenatbib]{mnras}

\usepackage{newtxtext,newtxmath}

\usepackage[T1]{fontenc}
\usepackage{ae,aecompl}
\usepackage{soul}

\usepackage{graphicx}	
\usepackage{amsmath}	
\usepackage{amssymb}	
\usepackage[usenames]{xcolor}
\usepackage{natbib}
\usepackage{hyperref}
\usepackage{xfrac}

\def\equationautorefname~#1\null{equation~(#1)}

\DeclareMathAlphabet{\mathpzc}{OT1}{pzc}{m}{it}\definecolor{purple}{RGB}{160,32,240}

\newcommand{\rev}[1]{\textcolor{black}{#1}}
\newcommand{\Msun}{M_{\odot}}
\newcommand{\Mstar}{M_{\star}}
\newcommand{\Mearth}{M_{\oplus}}

\newcommand{\Rearth}{R_{\oplus}}

\newcommand{\ta}{t_a}
\newcommand{\te}{t_e}
\newcommand{\tao}{t_{a,1}}
\newcommand{\teo}{t_{e,1}}
\newcommand{\tat}{t_{a,2}}
\newcommand{\tet}{t_{e,2}}
\newcommand{\Sigo}{\Sigma_0}

\title[Resonant Sub-Neptunes]{Sub-Neptune Formation: The View from Resonant Planets}

\author[Choksi \& Chiang]{Nick Choksi$^{1 \href{https://orcid.org/0000-0003-0690-1056}{\includegraphics[scale=0.4]{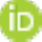}}}$\thanks{E-mail: nchoksi@berkeley.edu} and
Eugene Chiang$^{1,2\href{https://orcid.org/0000-0002-6246-2310}{\includegraphics[scale=0.4]{repulsion/orcid.pdf}}}$
\\
$^{1}$Astronomy Department, Theoretical Astrophysics Center, and Center for Integrative Planetary Science, University of California\\
\hspace{0.015in} Berkeley, Berkeley, CA 94720, USA\\
$^{2}$Department of Earth and Planetary Science, University of California, Berkeley, CA 94720, USA
}

\date{Released \today}

\pubyear{2020}

\begin{document}
\label{firstpage}
\pagerange{\pageref{firstpage}--\pageref{lastpage}}
\maketitle

\begin{abstract}
  The orbital period ratios of neighbouring sub-Neptunes are
  distributed asymmetrically near first-order resonances. There are
  deficits of systems---``troughs'' in the period ratio
  histogram---just short of commensurability, and
  excesses---``peaks''---just wide of it.  We reproduce quantitatively the strongest
  peak-trough asymmetries, near the 3:2 and 2:1 resonances, using dissipative
  interactions between planets and their natal discs. Disc
  eccentricity damping captures bodies into resonance and clears the
  trough, and when combined with disc-driven convergent migration,
  draws planets initially wide of commensurability into the peak. The
  migration implied by the magnitude of the peak is
  modest; reductions in orbital period are $\sim$10\%, supporting the
  view that sub-Neptunes complete their formation more-or-less in
  situ. Once captured into resonance, sub-Neptunes of typical mass
  $\sim$$5$--$15 \Mearth$ stay captured (contrary to an earlier claim),
  as they are immune to the overstability that afflicts lower mass
  planets.  Driving the limited, short-scale migration is a gas disc
  depleted in mass 
  relative to a solar-composition disc by 3--5 orders of
  magnitude. Such gas-poor but not gas-empty environments are
  quantitatively consistent with sub-Neptune core formation by giant
  impacts (and not, e.g., pebble accretion).  While disc-planet
  interactions at the close 
  of the planet formation era adequately explain the 3:2 and 2:1 asymmetries at 
  periods $\gtrsim$ $5$--$15$ days, 
  subsequent modification by stellar tides appears necessary at shorter
  periods, particularly for the 2:1.
\end{abstract}



\section{Introduction}
\label{sec:Intro}

\begin{figure*}
\includegraphics[width=\textwidth]{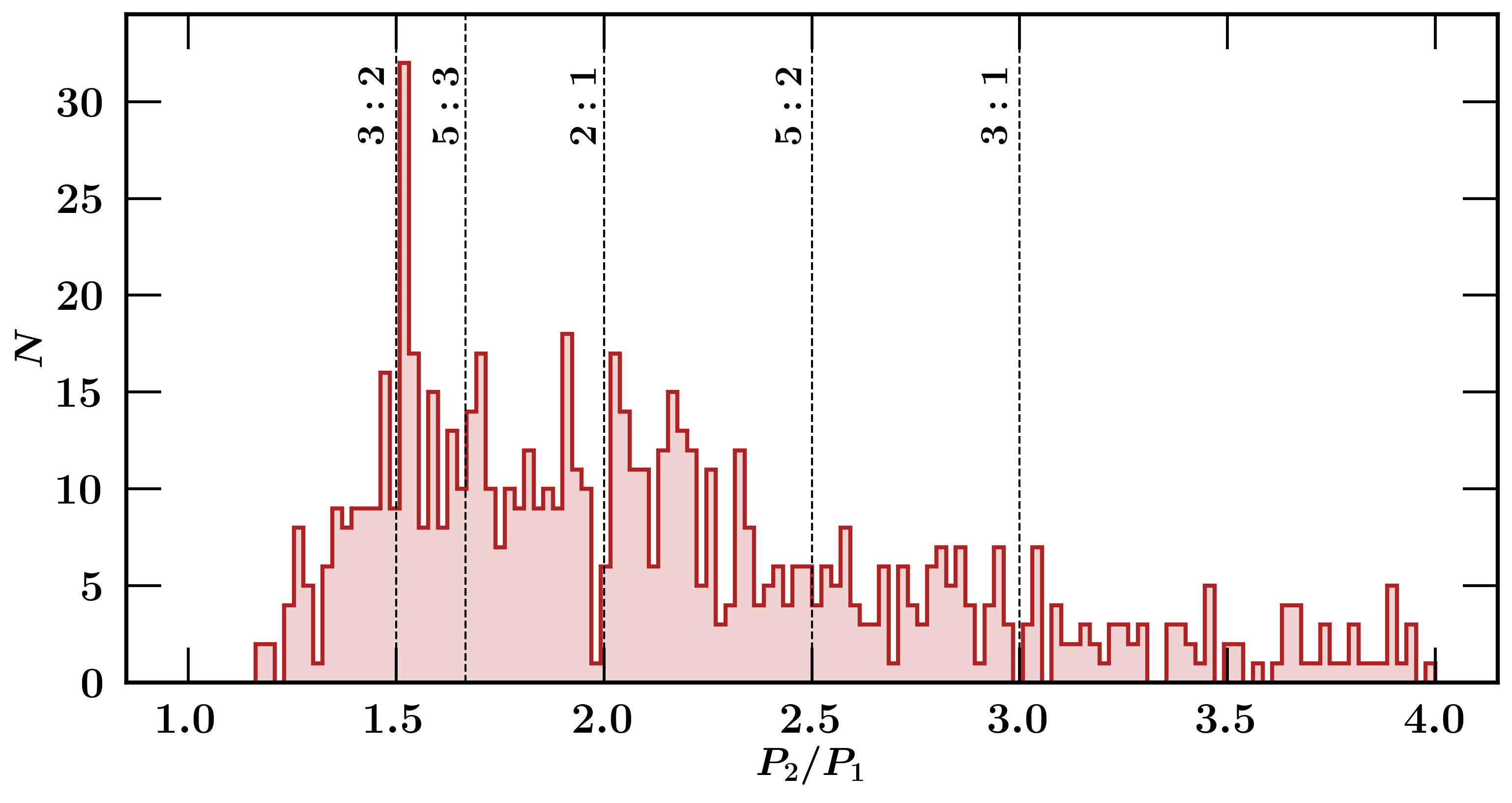}
\caption{Period ratios $P_2/P_1$ for all pairs of sub-Neptunes (with radii $< 4 \Rearth$) in the NASA Exoplanet Archive as of
August 1, 2019 \rev{($N$ is the number of systems
in a bin)}. Most sub-Neptunes are not in low-order
resonances, but there are excesses of systems (``peaks'')
just wide of the 3:2 and 2:1 commensurabilities,
and corresponding deficits (``troughs'') just short of
these resonances. See Appendix \ref{sec:app} for how these data separate by host star spectral type.}
  \label{fig:period_ratio_distributions}
\end{figure*}

As revealed by the \textit{Kepler} mission, sub-Neptunes (planets
with radii $\lesssim 4\Rearth$) are a dominant demographic,
orbiting an order-unity fraction of all FGKM stars with periods
less than a year \citep[e.g.,][]{fressin_etal_2013, dressing_charbonneau_2015, petigura_etal_2018,
zhu_etal_2018}.
And where there is one sub-Neptune orbiting a star, there is frequently at least
another \citep[e.g.,][]{zhu_etal_2018,sandford_etal_2019}.

\begin{figure*}
\vspace{-1.5cm}
\includegraphics[width=0.913\textwidth]{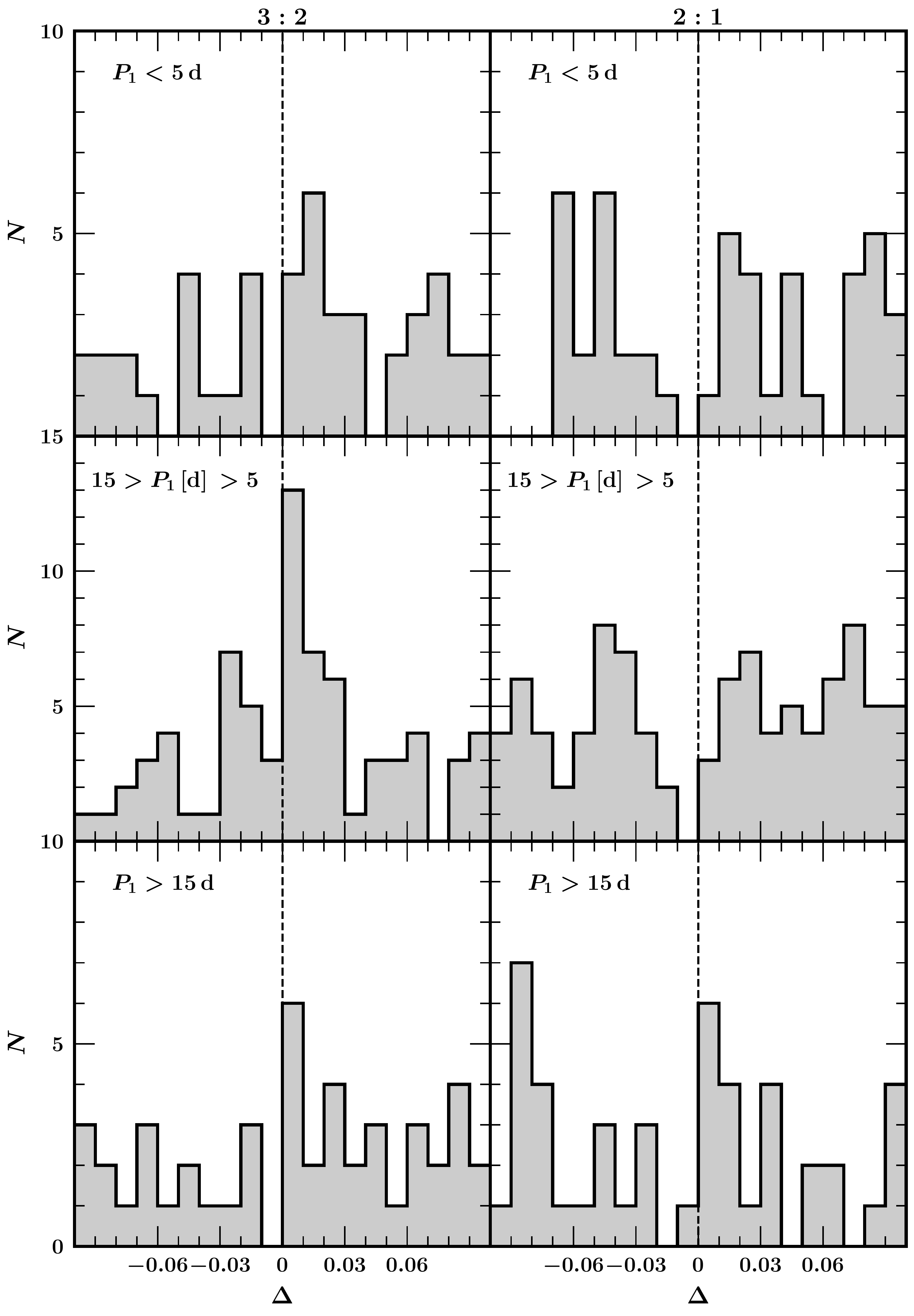}
\vspace{-.45cm}
\caption{Distribution of $\Delta$, the fractional separation from nominal resonance (equation \ref{eqn:delta_def}), for observed systems near the 3:2 and 2:1 commensurabilities \rev{($N$ is the number of systems in a bin)}. This figure is modeled after fig.~2 of \protect\cite{delisle_laskar_2014} 
who argued that the excess of systems just wide of 
resonance---what we call the ``peak''---diminishes
at large period ($P_1 > 15$ days),
apparently implicating tidal interactions
with the star which weaken rapidly with increasing
orbital distance. However, their figure 
employs a bin that is centered at $\Delta = 0$ and 
therefore mixes  $\Delta <0$ systems with
$\Delta >0$ systems, 
ignoring their qualitatively different dynamics.
Correcting the bin boundaries
recovers the peak and also its associated trough
at all periods, suggesting that tidal effects are not
sufficient to explain the asymmetry, especially at large period. Tides might still have a role to play in 
shifting the peak to larger $\Delta$ at the shortest
periods; this trend is stronger for the
2:1 than for the 3:2, as shown 
further in \autoref{fig:cdf}.}
  \label{fig:delta_psplit}
\end{figure*}

\cite{lissauer_etal_2011} and 
\cite{fabrycky_etal_2014}
measured the
period ratios $P_2/P_1$ of neighboring pairs of planets
(the subscript 1 denoting the inner member of the pair, and 2 the outer).  \autoref{fig:period_ratio_distributions}
presents an updated measurement of this period ratio distribution using the NASA
Exoplanet Archive.
For the most part, the distribution
of $P_2/P_1$ is broadly distributed between
$\sim$1.2 and 4, 
the lower bound marking the boundary
of dynamical stability (excepting planets
in 1:1 resonance, so far undetected).
Superposed on this continuum are excess numbers 
of planet pairs situated just wide of the 3:2 and 2:1 mean motion
commensurabilities. That is,
when populations are binned in $P_2/P_1$,
the bins situated just a percent or so larger than
3/2 or 2/1 
contain significantly more systems than
neighboring bins---there are resonant
``peaks'' in the histogram.
Accompanying these
peaks are ``troughs''---deficits of planet pairs
with period ratios a percent or so smaller than 3/2 or 2/1.
Similar substructure might also be present near the second-order 5:3 and 3:1
commensurabilities \citep[see][]{xu_lai_2017}. So far as we can tell, these period ratio
asymmetries are common to both FGK and M host stars (see Appendix \ref{sec:app}).

We use the dimensionless parameter
\begin{equation}
    \Delta \equiv \frac{q}{q+1}\frac{P_2}{P_1} - 1 
    \label{eqn:delta_def}
\end{equation}
to measure the deviation of the (instantaneous) period
ratio away from a first-order $(q+1)$:$q$ commensurability. The condition
$\Delta = 0$ is sometimes called ``nominal resonance", 
a condition not necessarily equivalent to the pair actually being ``in resonance''
or ``resonantly locked''; the latter terms imply that 
one or more resonant arguments librate
(i.e., one or more linear combinations of orbital longitudes oscillate about fixed points; e.g., \citealt{murray_dermott_1999}). 
The peak-trough asymmetry is an excess of planet pairs at $\Delta \sim 0.01$, and a deficit of
pairs at $\Delta \sim -0.01$, for $q=1$ and $q=2$.

The preference of resonant systems for $\Delta > 0$ can
be seen in the circular restricted planar three-body
problem. An inner test particle near a $(q+1)$:$q$ 
resonance with an outer planet of mass $\mu'$
relative to the central
star obeys the following equations, written here
to leading order in the test particle eccentricity and
to order-of-magnitude accuracy:
\begin{align}
\dot{n} &\sim \mu' e n^2 \sin \phi \\
\dot{e} &\sim \mu' n \sin \phi \\
\dot{\phi} &\sim (q+1)n' - qn + \frac{n\mu'}{e} \cos \phi
\end{align}
where 
$\phi = (q+1)\lambda' - q\lambda - \varpi$
is the resonant argument,
$\lambda$, $\varpi$, $e$, and $n \equiv 2\pi/P$ are the mean longitude,
longitude of periapse, eccentricity, and mean motion
of the test particle, and primed quantities
refer to the outer perturber on a fixed circular orbit. For an inner test
particle locked in resonance, $\phi$ librates
about the fixed point $\phi_0 = 0$; if the particle
resides at the fixed point with zero libration, $\dot{\phi} = 0$ and
\begin{align} \label{eq:simple_wedge}
(q+1)n' - qn \sim - \frac{n\mu'}{e} < 0
\end{align}
from which $\Delta > 0$ follows.
In other words, the inner test particle must speed
up its mean motion (relative to nominal resonance) if it is to repeatedly reach conjunction at
periapse in the face of apsidal regression; the smaller
the eccentricity, the faster the regression, and the more different the particle and perturber mean motions
have to be.
The same conclusion holds for the case of
an outer test particle 
resonantly locked with an interior perturber.
To be locked in a $(q+1)$:$q$ resonance with zero libration is actually to be
at a period ratio slighter greater than $(q+1)$:$q$,
in the absence of external sources of precession.

When resonantly locked planets have their orbital eccentricities
damped by an external agent, they are wedged farther
apart in semimajor axis 
($\dot{\Delta} > 0$;
\citealt{papaloizou_terquem_2010, lithwick_wu_2012}; \citealt{batygin_morbidelli_2013}). 
This ``resonant repulsion'' can be seen
in equation (\ref{eq:simple_wedge}),
whose right-hand side becomes more negative as $e$ decreases.
When eccentricites are damped following a fixed
time constant, $\Delta \propto t^{1/3}$ asymptotically.
Resonant repulsion has been proposed as a mechanism
to transport systems out of the trough at negative $\Delta$
and into the peak at positive $\Delta$.

\begin{figure}
\vspace{-1.5cm}
\includegraphics[width=0.95\columnwidth]{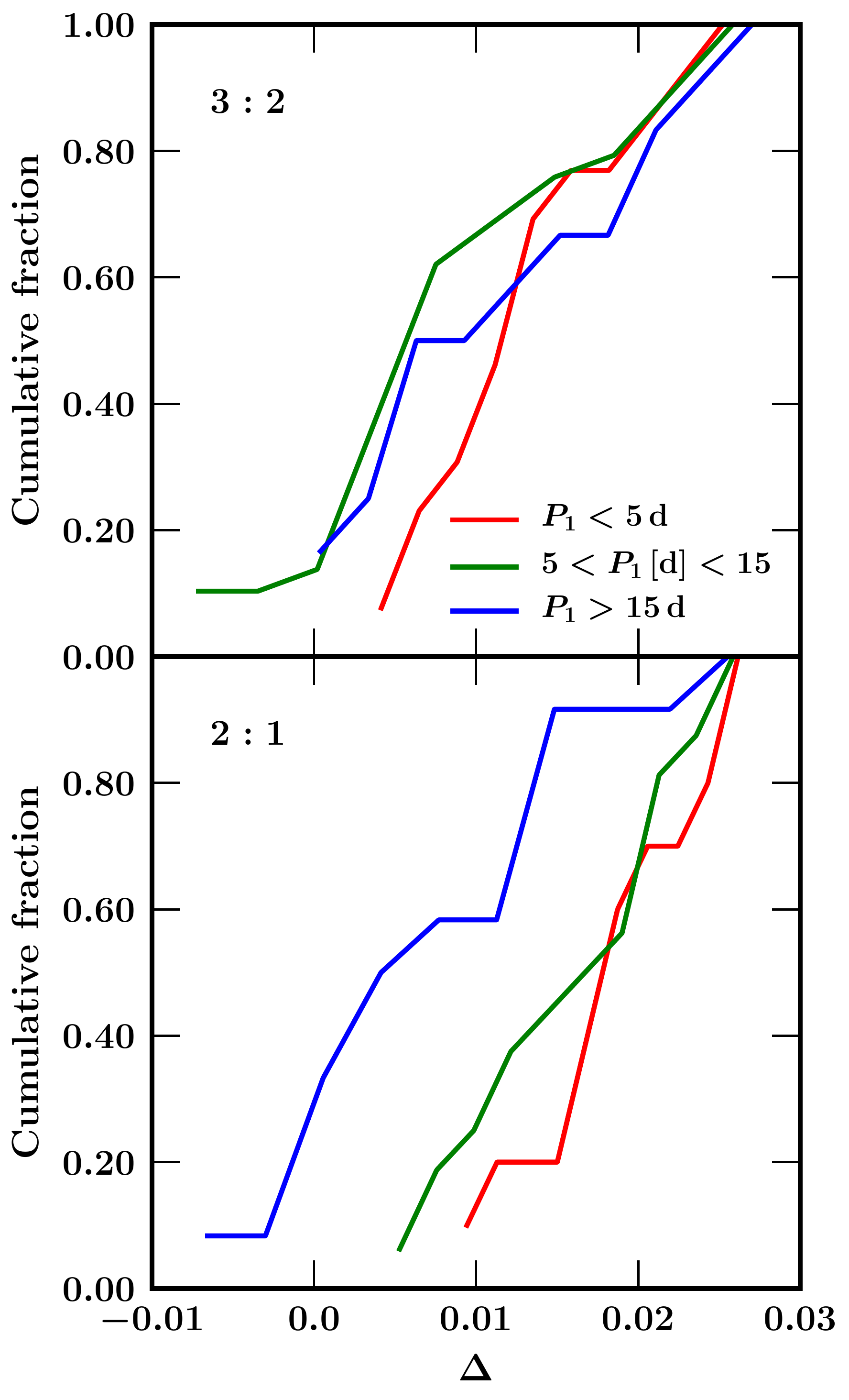}
\caption{Similar to \autoref{fig:delta_psplit}, but now plotting the cumulative distribution function,
which is not sensitive to choice of bins. This figure
confirms the trend reported by \citet[][their fig.~3]{delisle_laskar_2014} that planet
pairs situated closer
to the host star have larger $\Delta$ than those
situated farther away, and shows further that
this behaviour is stronger for the 2:1 resonance
than for the 3:2.}
  \label{fig:cdf}
\end{figure}

One way to damp eccentricities and drive repulsion is by
dissipating the eccentricity tide
raised on planets by their host stars. 
However, on the face of it, the tidal dissipation
rates required to reproduce the observed
$\Delta$-distribution within the system age
are too 
large compared to dissipation rates
inferred from Solar System planets (\citealt{lee_etal_2013, silburt_rein_2015}; but see \autoref{sec:summary_discussion}
where we discuss the proposal by \citealt{millholland_laughlin_2019} that dissipation can be provided by
obliquity tides). Furthermore, if tides, whose strength diminishes
rapidly with increasing distance from the host star, were the sole 
driver of resonant repulsion, the peak-trough asymmetry should become less pronounced at longer orbital periods.
From \autoref{fig:delta_psplit} we 
are hard pressed to say this is the case, 
as we can still make out the peak and the trough at periods
$\gtrsim 15$ days (cf.~\citealt{delisle_laskar_2014} who claimed otherwise, 
using a non-optimal binning scheme for their fig.~2;
see the caption to our \autoref{fig:delta_psplit}).
Tides might still have a role to play insofar as
the peak appears to shift to larger positive $\Delta$
with decreasing period (\citealt{delisle_laskar_2014}, their fig.~3);
our \autoref{fig:cdf} shows that this trend applies more to the 2:1 than to the 3:2. Our interpretation of these various
mixed (and low signal-to-noise) messages is that
tidal interactions with the star
may have shaped the period ratio asymmetry 
at the shortest periods, but 
may not be the whole story, especially at long periods.

Another way to damp eccentricities is by torques exerted
on planets by their parent gas discs, during
the planet formation era \citep{goldreich_tremaine_1980,artymowicz_1993, cresswell_etal_2007}.
In addition to damping planet eccentricities, discs
also change planet semimajor axes
\citep{goldreich_tremaine_1980}, i.e., they drive
orbital migration, typically
toward the star 
\citep{ward_1997, kley_nelson_2012}.
Planets that migrate convergently (toward smaller
$P_2/P_1$) can become captured into mean-motion resonance.
Whether the resonance is stable in the face
of continued migration and eccentricity damping depends
on the planet masses \citep{meyer_wisdom_2008, goldreich_schlichting_2014, deck_batygin_2015}. Sufficiently high planet masses
lead to permanent capture, with eccentricity pumping by resonant migration balancing eccentricity damping by the disc 
\citep[e.g.,][]{lee_peale_2002}. 
The equilibrium eccentricities so established
imply a positive equilibrium value for $\Delta$
(cf.~equation \ref{eq:simple_wedge}) 
that depends on planet-to-star mass ratios
and the relative rates at which the disc
drives eccentricity and semimajor axis changes
\citep{terquem_papaloizou_2019}.

In this paper we ask whether disc-planet interactions
can reproduce the observed peak-trough features in the
$\Delta$-distribution near first-order
resonances. 
We seek to use the observed
period ratio distribution to constrain the extent
to which sub-Neptunes migrated, a question
tied to how much gas was present in the parent disc
around the time these planets finished forming (i.e.,
completed their last doubling in mass).
On the one hand, the observation that most planets
neither lie near a period commensurability nor
pile up at short periods suggests the majority of systems
formed in situ \citep[e.g.,][]{lithwick_wu_2012, elee_chiang_2017, terquem_papaloizou_2019, macdonald_etal_2020},
consistent with formation models staged late in a disc's life,
when little gas remains to drive migration \citep[e.g.,][]{kominami_ida_2002, elee_chiang_2016, elee_etal_2018}.
On the other hand there are gas-rich scenarios
for sub-Neptune formation---pebble accretion falls in 
this category \citep[e.g.,][]{bitsch_etal_2019, lambrechts_etal_2019,rosenthal_murray-clay_2019}---where the many systems
caught into resonance by disc-driven migration must
eventually escape resonance, ostensibly because of
instabilities driven by disc eccentricity damping 
\citep{goldreich_schlichting_2014, deck_batygin_2015}
or chaos in high-multiplicity systems \citep{pu_wu_2015, izidoro_etal_2017, izidoro_etal_2019}. Our goal is to 
help decide the in-situ vs.~migration (gas-poor vs.~gas-rich disc)
debate for sub-Neptunes by quantifying the disc gas surface density
and the extent of planet-disc interaction
needed to reproduce the peak-trough
asymmetry revealed by ${\it Kepler}$.

We begin in \autoref{sec:methodology} by laying out
the equations of motion solved in this paper
for near-resonant, disc-driven pairs of planets.
In \autoref{sec:restricted} we review, for the special
case of the circular
restricted planar three-body problem, the behaviour of near-resonant test
particles whose semi-major axes and eccentricities are externally driven by a disc.
There we survey the various possible evolutions for $\Delta$.
\autoref{sec:unrestricted} describes
how these results are modified when the masses of both
planets are accounted for.
Our main contribution is in \autoref{sec:mc} 
where we carry out a population synthesis,
generating mock populations of planet pairs that evolve under
the influence of a disc, 
and comparing our calculated $\Delta$-distributions to the observed
$\Delta$-distribution to constrain disc properties. 
In \autoref{sec:summary_discussion}
we place our results in the context of our understanding of planet
formation and identify areas for future work.

By design our paper studies planet-disc interactions
and does not model stellar tidal interactions.
Most of our calculations (all those
in Sections \ref{sec:unrestricted}--\ref{sec:mc})
will be for the 3:2 resonance, which exhibits the strongest
peak-trough asymmetry and the one least sensitive
to distance from the host star (Figures 
\ref{fig:period_ratio_distributions}--\ref{fig:cdf}).
Our hypothesis is that 3:2 systems are least
impacted by tides. 
We bring the 2:1 resonance back into consideration
in Section \ref{sec:summary_discussion}.
There we assess the extent to which disc-planet interactions, which establish a baseline for the
peak-trough asymmetry, need to be abetted by tides.

\section{Equations of Motion}
\label{sec:methodology}
 
To leading order in eccentricity, 
two planets of mass $m_1$ and $m_2$
orbiting a star of mass $\Mstar$
near a $(q+1)$:$q$ mean motion resonance 
obey the following  
coupled ordinary differential equations
for their mean motions $n$,
eccentricities $e$,
and resonant arguments $\phi$ 
(subscript 1 for the inner planet and 2 for the outer planet; 
e.g., \citealt{terquem_papaloizou_2019}):

\begin{align}
    \dot{n}_1  &= -3qn_1^2\frac{\alpha m_2}{\Mstar}\left(e_1 f_1 \sin\phi_1 + e_2 f_2 \sin\phi_2 \right) \nonumber \\ &+ \frac{3n_1}{2\tao} + \frac{pn_1 e_1^2}{\teo}  \label{eqn:n1dot}  \\ 
    \dot{n}_2 &= 3(q+1)n_2^2 \frac{m_1}{\Mstar}\left(e_1 f_1 \sin\phi_1 + e_2 f_{2}\sin\phi_2\right) \nonumber \\ &+ \frac{3n_2}{2\tat} + \frac{pn_2 e_2^2}{\tet} \label{eqn:n2dot}   \\
    \dot{e}_1 &= -n_1\frac{\alpha m_2}{\Mstar}f_1\sin\phi_1 - \frac{e_1}{\teo} \label{eqn:e1dot}    \\ 
    \dot{e}_2 &= -n_2\frac{m_1}{\Mstar}f_2\sin\phi_2 - \frac{e_2}{\tet} \label{eqn:e2dot}   \\ 
    \dot{\phi}_1 &= (q+1)n_2 - qn_1 - n_1\frac{\alpha m_2}{\Mstar}\frac{1}{e_1}f_1\cos\phi_1 \label{eqn:phi1dot}    \\ 
    \dot{\phi}_2  &= (q+1)n_2 - qn_1 - n_2\frac{m_1}{\Mstar}\frac{1}{e_2}f_2\cos\phi_2 \,. \label{eqn:phi2dot}
\end{align}
The coefficients $f_1$ and $f_2$ are given in terms of
the ratio of semimajor axes $\alpha \equiv a_1/a_2$
and Laplace coefficients:
\begin{align}
    f_1 = -\frac{1}{2}\left[2(q+1) + \alpha\frac{d}{d\alpha}\right]b_{1/2}^{q+1}(\alpha) \\
    f_2 = \frac{1}{2}\left[2q + 1 + \alpha\frac{d}{d\alpha}\right]b_{1/2}^{q}(\alpha) - 2\alpha \delta_{q,1} \\ 
    b_{1/2}^{j}(\alpha) = \frac{1}{\pi}\int_0^{2\pi}\frac{\cos(j\psi)}{\left(1 - 2\alpha\cos\psi + \alpha^2\right)^{1/2}}d\psi
\end{align}
where $\delta_{q,1}$ is the Kronecker $\delta$.

We focus on the case $q=2$, i.e., the $n_1$:$n_2$ = 3:2 resonance for which the observed peak-trough asymmetry is strongest and least
sensitive to orbital distance (read: least affected
by stellar tidal interactions; Figures \ref{fig:period_ratio_distributions}--\ref{fig:cdf}). 
We hold fixed $f_1 = -2.025$ and $f_2 = 2.484$, the values
appropriate for $\alpha = a_1/a_2 = (2/3)^{2/3}$ at nominal resonance. In reality, $\alpha$ varies with time,
but by amounts too small for the resultant
changes to $f_1$ and $f_2$ to matter.

To the resonant interaction terms (those depending on $\phi$
in equations \ref{eqn:n1dot}--\ref{eqn:e2dot})
we have added terms for semi-major axis and eccentricity damping
by an external agent---in this paper, the disc---parameterized 
by the timescales $\ta$ and $\te$. 
The coefficient $p$ measures 
the extent to which eccentricity damping alone
(ignoring the resonant potential) produces semi-major axis changes.
If eccentricity damping alone conserved a planet's orbital angular
momentum, then $p=3$. Although disc torques
(first-order co-orbital Lindblad torques in the case of eccentricity damping; e.g., \citealt{duffell_chiang_2015} and references
therein)
generally do not conserve the planet's angular momentum, the
relevant value of $p$ may differ from 3 only by an order-unity factor. 
Moreover, both resonant repulsion \citep{lithwick_wu_2012}
and resonant equilibria \citep{goldreich_schlichting_2014, terquem_papaloizou_2019}
are not too sensitive to $p$ (which could even be 0).
For simplicity, and following previous work, 
we adopt $p=3$.
Note further that the effect of the disc
on apsidal precession has been neglected;
the last terms in equations (\ref{eqn:phi1dot}) and (\ref{eqn:phi2dot}) account only for precession due to the resonance.

\begin{figure}
\includegraphics[width=\columnwidth]{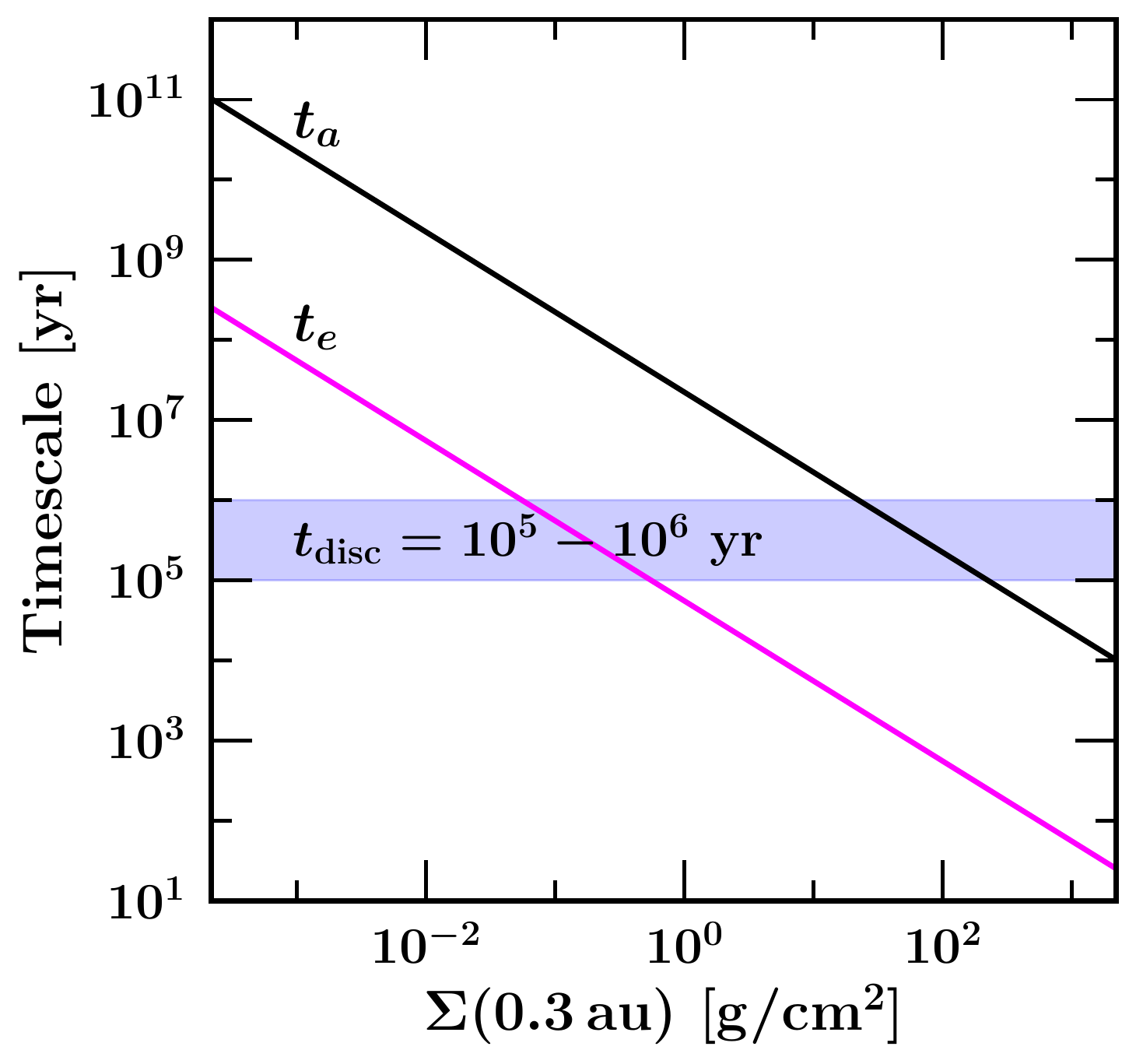}
\vspace{-0.5cm}
\caption{Characteristic semi-major axis and eccentricity damping
  times $\ta = a/|\dot{a}|$ and $\te = e/|\dot{e}|$ for a sub-Neptune
in a disc, as a function of the disc gas surface density $\Sigma$,
evaluated for a planet mass of $10\,\Mearth$ and an orbital radius of
$a = 0.3$ au. The shaded region 
indicates possible disc dispersal (e-folding) timescales. 
Eccentricity damping by
the disc is faster than semi-major axis damping (orbital migration)
by a factor of order $(a/h)^2$.}
  \label{fig:timescales}
\end{figure}

For the semi-major axis damping time $t_a$ we 
utilize the numerically calibrated value of 
\cite{kley_nelson_2012}:
\begin{equation}
    \ta =  \frac{m \sqrt{G\Mstar a}}{2|\Gamma|} 
    \label{eqn:ta}
\end{equation}
\begin{equation}
     \Gamma = -\left(1.36 + 0.62\beta_{\Sigma} + 0.43\beta_T\right) 
 \left(\frac{m}{\Mstar}\right)^2\left(\frac{h}{a}\right)^{-2}\Sigma a^4 \Omega^2 
\end{equation}
where $G$ is the gravitational constant, and
$\Sigma$, $h/a$, and $\Omega$ are the disc surface
density, aspect ratio, and Keplerian angular frequency
evaluated at the planet's semimajor axis $a$, respectively.
The variables $\beta_T \equiv -d\log T/d\log a$
and $\beta_\Sigma \equiv -d\log \Sigma/d\log a$ are
the power-law indices describing how temperature
and surface density vary with disc radius. We assume
$\beta_T = 3/7$ \citep{chiang_goldreich_1997} and set
\begin{equation}
h/a = 0.04 \left( \frac{a}{1\, {\rm au}} \right)^{2/7} \,.
\end{equation}
For most of our calculations we choose for simplicity
$\beta_\Sigma = 0$. A flat $\Sigma$ profile
yields nearly equal fractions of convergently
and divergently migrating planet pairs,
assuming $m_2$ and $m_1$ are drawn independently
from the same distribution. 
However, we
also experiment with $\beta_\Sigma$ up to 3/2
(the value appropriate to the minimum-mass
solar and extrasolar nebulas; \citealt{chiang_laughlin_2013}).
We assume $\Sigma$ 
decays exponentially with time:
\begin{equation} \label{eqn:sigo}
    \Sigma(a,t) = \Sigo \left( \frac{a}{1\,{\rm au}}
    \right)^{-\beta_\Sigma} \exp (-t/t_{\rm disc}) 
\end{equation}
with a nominal $t_{\rm disc} = 10^5$ yr, arguably appropriate for the innermost regions of discs where {\it Kepler} sub-Neptunes reside \citep[e.g.,][]{alexander_etal_2014}. 
The initial surface density normalization $\Sigo$ is a free parameter that we will fit to the observations (\autoref{sec:mc}).

The eccentricity damping timescale is given by
\begin{equation}
    \te = \left(\frac{\Mstar}{m}\right)\left(\frac{\Mstar}{\Sigma a^2}\right)\left(\frac{h}{a}\right)^4\Omega^{-1} 
    \label{eqn:te}
\end{equation}
(e.g., \citealt{kominami_ida_2002}). 
There are corrections to $\te$ that
grow with $e/(h/a)$ \citep{papaloizou_larwood_2000},
but these are less than order-unity
for the small eccentricities considered here
and are therefore omitted \citep[cf.][]{xu_etal_2018}. \autoref{fig:timescales} plots sample values of
$\ta$ and $\te$ as a function of $\Sigma$,
for $a = 0.3$ au and $m = 10 \,\Mearth$. 
For our disc parameters, the ratio $\ta/\te$
for a single planet 
varies from 200 to 750
as $a$ varies from 1 au to 0.1 au.

The migration we model is smooth and of varying
rates depending on the gas surface density. 
\cite{rein_2012} also study near-resonant planets torqued
by discs, but focus on stochastic migration in turbulent, gas-rich discs. Their model employs a fixed value for the ratio of damping timescales $\ta/\te = 10$ that appears underestimated by more than an order of magnitude.

Equations (\ref{eqn:n1dot})--(\ref{eqn:phi2dot})
are solved numerically for how the distance from period
commensurability $\Delta$ evolves for a pair of planets
embedded in a decaying disc. 
We carry out all numerical integrations using the \textsc{lsoda} package, enforcing a
fractional tolerance of $10^{-10}$ on the accuracy of our solutions. 
As a check on our calculations, we compared
them against analytic equilibrium solutions for $e_1$,
$e_2$, and $\Delta$ as 
derived by Terquem \& Papaloizou
(\citeyear{terquem_papaloizou_2019}; their equations 35, 36, and 49):\footnote{Our definition of $\Delta$ differs from that of \citet[][TP]{terquem_papaloizou_2019}:
$\Delta_{\rm TP} \equiv -q\Delta(\Delta + 1) \simeq -q\Delta$, where for the last equality we have used $\Delta \ll 1$, a condition valid everywhere in our paper.}
\begin{align}
   e_{\rm eq,1} &= \left\{ \frac{  \teo/\tat - \teo/\tao }{2\left(q+1\right) \left(1 + \frac{q}{q+1} \frac{m_1}{\alpha m_2} \right) \left[1 + \frac{m_1}{\alpha m_2} \left( \frac{q}{q+1} \right)^2 \left( \frac{f_2}{f_1} \right)^2 \frac{\teo}{\tet} \right] \label{eqn:tp1}
   } \right\}^{1/2}\\
    e_{\rm eq,2} &= e_{\rm eq,1} \left( \frac{m_1}{\alpha m_2} \right) \left( \frac{q}{q+1} \right) \left| \frac{f_2}{f_1} \right| \label{eqn:tp2}
    \\
\Delta_{\rm eq} &= \sqrt{-\mathcal{A}/\mathcal{B}} \nonumber \\
    \mathcal{A} &=  \frac{3}{q^2 \teo}\left(\frac{q}{q+1}\frac{m_1}{\alpha m_2} + 1 \right)\left(\frac{\alpha m_2}{\Mstar}\right)^2 \times \nonumber \\   &\left[(q+1)f_1^2 + \frac{q^2}{q+1}\frac{m_1}{\alpha m_2}f^2_2\frac{\teo}{\tet}\right] > 0 \nonumber  \\ 
    \mathcal{B} &= \frac{3}{2\tao}\left(1 - \frac{\tao}{\tat}\right) < 0 \,.
    \label{eqn:delta_dot} 
\end{align}
Application of equations (\ref{eqn:tp1})--(\ref{eqn:delta_dot}) is restricted to the case
where a pair of planets migrate convergently (so
$\mathcal{B} < 0$, either because (i) 
$\tao/\tat > 1$ for $\tao,\tat > 0$, 
(ii) $\tao < 0$ and $\tat > 0$, or (iii)
$\tao/\tat < 1$ for $\tao, \tat < 0$)
and locked in mutual resonance 
(at $\cos \phi_1 = 1$ and $\cos \phi_2 = -1$). Our numerical
solutions to the more general equations (\ref{eqn:n1dot})--(\ref{eqn:phi2dot})
do not make these assumptions.

\rev{Some additional notes on the validity of equations
(\ref{eqn:n1dot})--(\ref{eqn:phi2dot}): the terms depending
on $\phi$ reasonably describe the interaction potential
``near resonance'', meaning either when $\phi$ is librating
about a fixed value (in resonance), or when $\phi$ is 
circulating from $0$ to $2\pi$ outside and not
too far from resonance
(see, e.g., \citealt{murray_dermott_1999}, their figure 8.16, panels e and f).
Far from resonance, keeping these terms and neglecting other short-period terms is technically not accurate, but the error is of little consequence, as here all terms depending on mean longitudes time-average to zero anyway. For our model parameters, the planets are
never situated so far from resonance that their
forced eccentricities (forced by the resonance) are smaller than the forced eccentricities
from other short-period terms; the former are of order $\mu / \Delta$, where $\Delta < 0.1$ for
all calculations in this paper, while the latter are of order $\mu$ (\citealt{agol_etal_2005}, their section 5). These eccentricities, and our modeled eccentricities which by assumption start at values $< 0.1$
and decrease by disc damping, are all small enough
that a first-order expansion of the resonance
potential suffices.}

\section{The Restricted Problem}
\label{sec:restricted}
\begin{table*}
\centering
\begin{tabular}{|c|c|c|c|c|c|c|c|c|}
\hline
\\[-2mm]
Case & Test particle & $\mu$ & $|\ta|/\te$ & Migration &  $\phi_{\rm final}$ & Resonant repulsion? & $\Delta_{\rm final}$ & $e_{\rm final}$ \\  \hline \hline
1 & Inner & Any & $\infty$ & None &   0$^+$ & Yes & $\Delta_{\rm eq} = \infty$ &  0    \\  \hline 
2 & Inner & $> \mu_{\rm crit}$ & 100 & Convergent &   0$^+$ & Yes & $\Delta_{\rm eq}$ & $e_{\rm eq,1}$    \\  \hline
3 & Inner & $< \mu_{\rm crit}$ & 100 & Convergent &   $\pi^{-}$ & No & $\ll \Delta_{\rm initial}$ & 0      \\  \hline 
4 & Inner & Any & 100 & Divergent &    $0^{+}$ & No & $\gg \Delta_{\rm initial}$ & 0          \\  \hline 
5 & Outer & Any & $\infty$ & None &  $\pi^+$ & Yes & $\Delta_{\rm eq} = \infty$ & 0      \\  \hline
6 & Outer & Any & 100 & Convergent & $\pi^+$ & Yes & $\Delta_{\rm eq}$ &  $e_{\rm eq, 2}$  \\     \hline 
7 & Outer & Any & 100 & Divergent &  $\pi^{+}$ & No & $\gg \Delta_{\rm initial}$ &  0  \\  \hline
\end{tabular}
\caption{Summary of the behaviour of a test particle in the restricted planar circular three-body problem near the 3:2 resonance, when the test particle's eccentricity is damped and when its semi-major axis is driven either outward or inward. Columns specify: (1) case number, (2) 
whether the test particle resides interior or exterior to the 
perturber (``Inner'' vs.~``Outer''), (3) the perturber-to-star mass
  ratio, with
the value $\mu_{\rm crit}$ (above which resonance
capture is permanent and below which it is not) given by equation (28) of \citet{goldreich_schlichting_2014}, 
(4) the ratio of the externally imposed e-folding timescales for the test particle semi-major axis and eccentricity, where
$\infty$ denotes infinite $\ta$ and finite $\te$, (5) whether
migration is convergent or divergent (as controlled by the sign of
$\ta$) or is not imposed, (6) the late-time value of the test
particle's resonant angle $\phi$, with superscripts ``+'' and ``-''
denoting values slightly greater or less than the listed number, (7)
whether or not the planet-particle pair exhibits resonant repulsion,
(8) the late-time value of the pair's distance from nominal resonance
$\Delta$, 
with equilibrium values $\Delta_{\rm eq}$ given by equations
(\ref{eqn:delta_eq}) and (\ref{eqn:delta_eq2}), (9) the late-time
value of the eccentricity, with equilibrium values $e_{\rm eq}$ given by
equations (\ref{eqn:e_eq}) and (\ref{eqn:eeq2}). Cases 1, 2, 5, and 6 can lead to permanent capture into resonance; cases 3, 4, and 7 cannot.}
  \label{tab:cases_gs}
\end{table*}

\begin{figure}
\includegraphics[width=0.96\columnwidth]{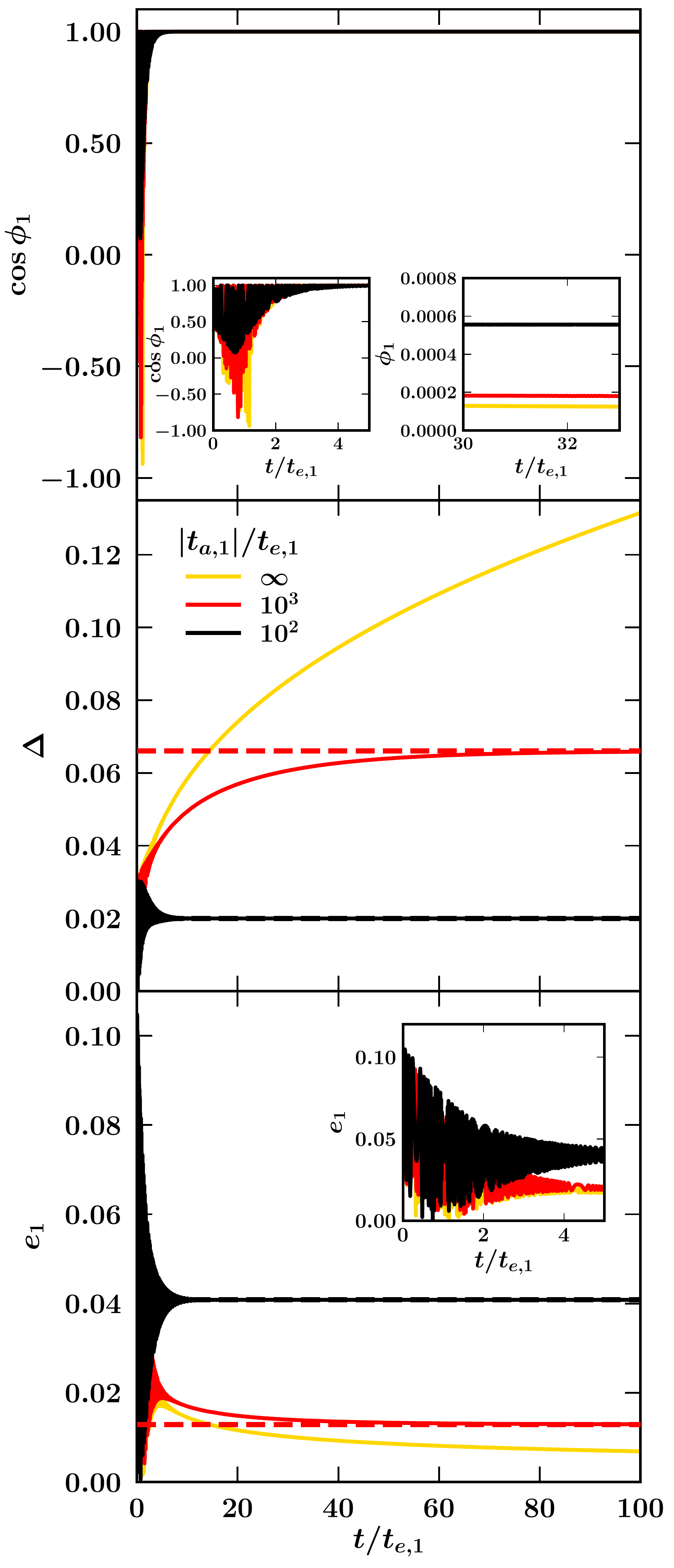}
\caption{Evolution of an inner test particle whose semi-major
axis and eccentricity are externally driven, near the 3:2 resonance with an outer massive
perturber having $\mu = 10^{-3}$ on a fixed circular orbit. The test particle
is captured into resonance ($\phi_1$ locks
to a stable point slightly greater than 0; top panel) either by 
eccentricity damping operating alone (yellow curves),
or a combination of eccentricity
damping and convergent migration (red and black curves).
In the former case, the test particle eccentricity $e_1$ asymptotes to
zero (bottom panel) and the separation $\Delta$ from
nominal resonance increases
as $t^{1/3}$
(middle panel; \citealt{lithwick_wu_2012}; \citealt{batygin_morbidelli_2013}).
When convergent migration is added, resonant amplification of eccentricity
balances disc eccentricity damping to yield
an equilibrium eccentricity $e_{\rm eq,1}$
and an equilibrium separation $\Delta_{\rm
  eq}$; their values calculated analytically 
  from equations (\ref{eqn:tp1}) and (\ref{eqn:delta_dot}) are shown
as dashed lines 
and agree with our numerical results.}
  \label{fig:var_tae}
\end{figure}

\begin{figure}
\includegraphics[width=\columnwidth]{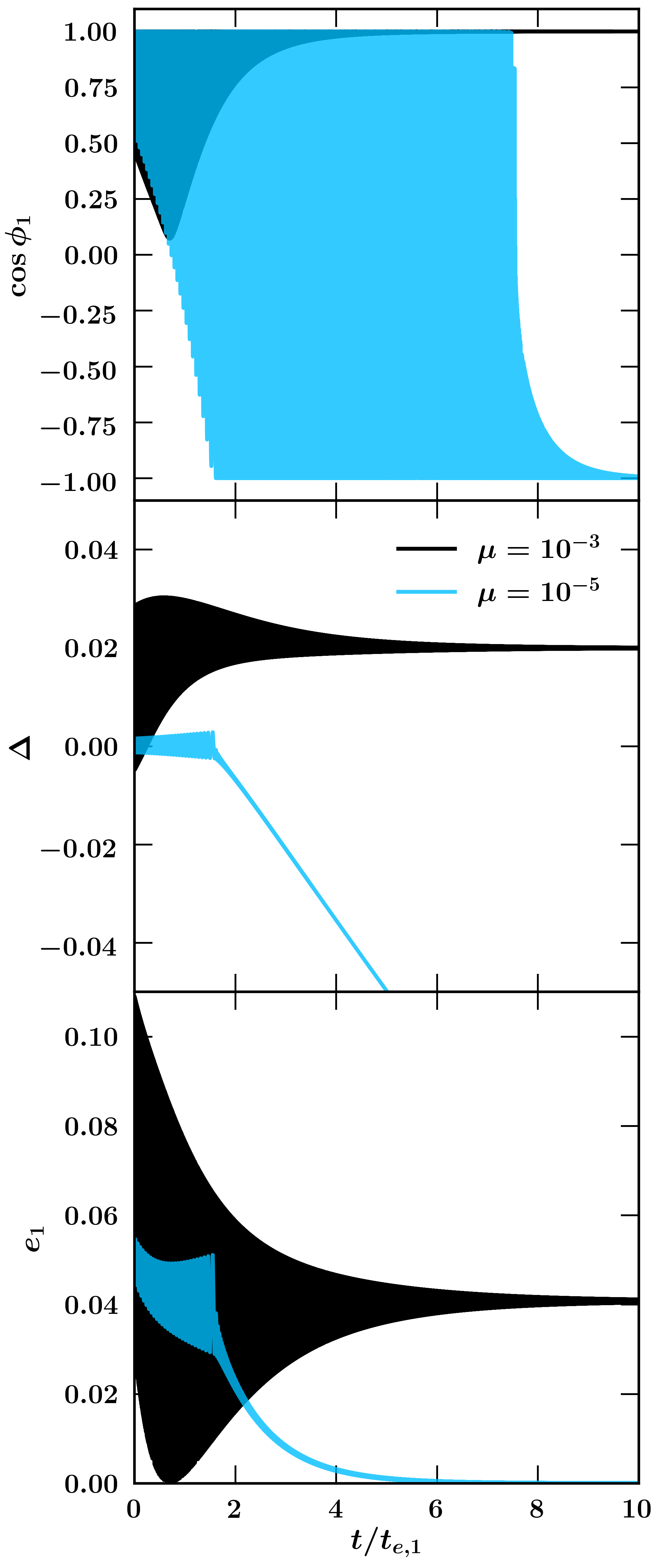}
\caption{Same as \autoref{fig:var_tae}
but for fixed $|\tao|/\teo = 100$ and two 
perturber-to-star mass ratios $\mu$. 
For low $\mu$ the inner test particle experiences
overstable librations and eventually escapes
resonance (blue curve).}
  \label{fig:gs_escape}
\end{figure}

To gain intuition and connect to previous work, we first explore the restricted problem where one of the planets is replaced with a test particle, while the other planet of non-zero mass
$m \equiv \mu M_\star$ is kept on a fixed circular
orbit (in this section we forgo the primed vs.~unprimed
notation). The damping timescales $\ta$ and $\te$ refer here
to the test particle, and are held constant
in a given integration for simplicity (they do not refer
to a depleting disc per se). 
We explore both the cases of an inner test particle
($m_1, e_2 \rightarrow 0$ and $\tat, \tet \rightarrow \infty$ in equations \ref{eqn:n1dot}-\ref{eqn:phi2dot})
and outer test particle ($m_2, e_1 \rightarrow 0$, $\tao, \teo
\rightarrow \infty$), as well as both convergent and divergent
migration (controlled by the sign of $\ta$ which we allow here to be negative). 
For the integrations reported in this section,
initial conditions are as follows: 
$\Delta_{\rm initial} = 0$, test particle $\phi_{\rm initial} = 1$ rad (away from the
fixed points of the resonance near 0 and $\pi$), and test particle $e_{\rm initial} = 0.05$. 
Other initializations give qualitatively similar outcomes.

We begin with the case where an inner test particle is subject only to
eccentricity damping ($\teo$ finite, $\tao = \infty$). 
From \autoref{fig:var_tae}, made for an outer perturber of mass 
$\mu = 10^{-3}$,
we see the test particle
lock into resonance on a timescale of order $\teo$ (top panel, left inset).
The resonant angle $\phi_1$ settles to a small positive
value (top panel, right inset). 
This small offset in $\phi_1$ away from 0 (the
dissipationless equilibrium point) arises because
eccentricity damping 
accelerates apsidal regression 
($\dot{\varpi}_1 \propto -1/e_1$), causing
conjunctions to occur
just after periapse. 
Such conjunctions remove
angular momentum
from the test particle (e.g., \citealt{peale_1986}),
driving  
it away indefinitely according to $\Delta \propto t^{1/3}$ (middle panel, yellow curve). \citet{lithwick_wu_2012} term this behaviour resonant repulsion---the bodies, locked
in resonance, are repelled farther apart. 

\autoref{fig:var_tae} also shows that
for convergent migration at finite $\tao < 0$
($|\tao|$ is allowed to
vary from $10^2\teo$ to $10^3\teo$) 
the planet-particle pairs do not wedge apart for all time 
but reach an equilibrium separation
$\Delta_{\rm eq} > 0$, i.e., they reach an
equilibrium wide of resonance.
From \autoref{eqn:delta_dot},
\begin{align}
    \Delta_{\rm eq} = -\frac{\alpha\mu f_1}{q}\sqrt{2(q+1)\frac{|\tao|}{\teo}} > 0
    \label{eqn:delta_eq}
\end{align}
which agrees with our
numerical results (middle panel). 
This equilibrium $\Delta_{\rm eq}$
corresponds to an equilibrium
eccentricity $e_{\rm eq,1}$ (given by equation A1 of
\citealt{goldreich_schlichting_2014},
or our equation \ref{eqn:tp1} taken from
\citealt{terquem_papaloizou_2019}; see also equation
\ref{eq:simple_wedge}).
The equilibrium eccentricity
reflects the balance between eccentricity
pumping by resonant migration (driven by $\tao$)
and eccentricity damping by the disc ($\teo$):
\begin{align}
        e_{\rm eq,1} = 
        \sqrt{\frac{1}{2(q+1)} \frac{\teo}{|\tao|}}
    \label{eqn:e_eq}
\end{align}
which also matches our numerical results (bottom panel). 

The value of $\mu = 10^{-3}$ in \autoref{fig:var_tae}
was chosen to exceed $\mu_{\rm crit} \propto (\teo/|\tao|)^{3/2}$,
the value above which the resonance is stable to eccentricity damping and below
which it is not; we find using equation (28) of 
\citet{goldreich_schlichting_2014} that
$\mu_{\rm crit} \simeq 6 \times 10^{-5}$ for
$|\tao|/\teo = 10^2$. 
In \autoref{fig:gs_escape} we verify that for a lower
perturber mass, 
$\mu = 10^{-5}$, 
the test particle
eventually escapes resonance after a few $\teo$,
after which its eccentricity decays exponentially to zero,
and $\Delta$ becomes increasingly negative
because of the imposed convergent migration
(i.e., equation \ref{eqn:n1dot} reduces to
$\dot{n}_1 = 3n_1/(2\tao)$).

The behaviours discussed so far for the inner test particle 
are summarized as cases 1, 2, and
3 in \autoref{tab:cases_gs}. Divergent migration for the inner
test particle, case 4, does not lead to permanent resonance capture \citep[e.g.,][]{murray_dermott_1999} and simply causes
$\Delta$ to increase on timescale $\tao$. 
\autoref{tab:cases_gs} also provides entries
for an outer test particle.
An outer test particle behaves similarly to an inner
test particle, except that all perturber
masses can permanently
capture an outer test particle 
when migration is convergent (contrast cases 2 and 3 for
the inner test particle with the single case 6
for an outer test particle; see also section 2.2.2
of \citealt{deck_batygin_2015}).
When an outer test particle migrates convergently and is resonantly captured,
the equilibrium separation and eccentricity are given by the appropriate limits of equations (\ref{eqn:tp1})--(\ref{eqn:delta_dot}):
\begin{align} \label{eqn:delta_eq2}
 \Delta_{\rm eq} = \frac{\mu f_2 q}{q+1}\sqrt{\frac{2}{q} \frac{\tat}{\tet}} > 0\\
    e_{\rm eq,2} = \sqrt{\frac{1}{2q}\frac{\tet}{\tat}} \,. \label{eqn:eeq2}
\end{align}

\section{The unrestricted problem}
\label{sec:unrestricted}

\begin{figure}
\includegraphics[width=0.9\columnwidth]{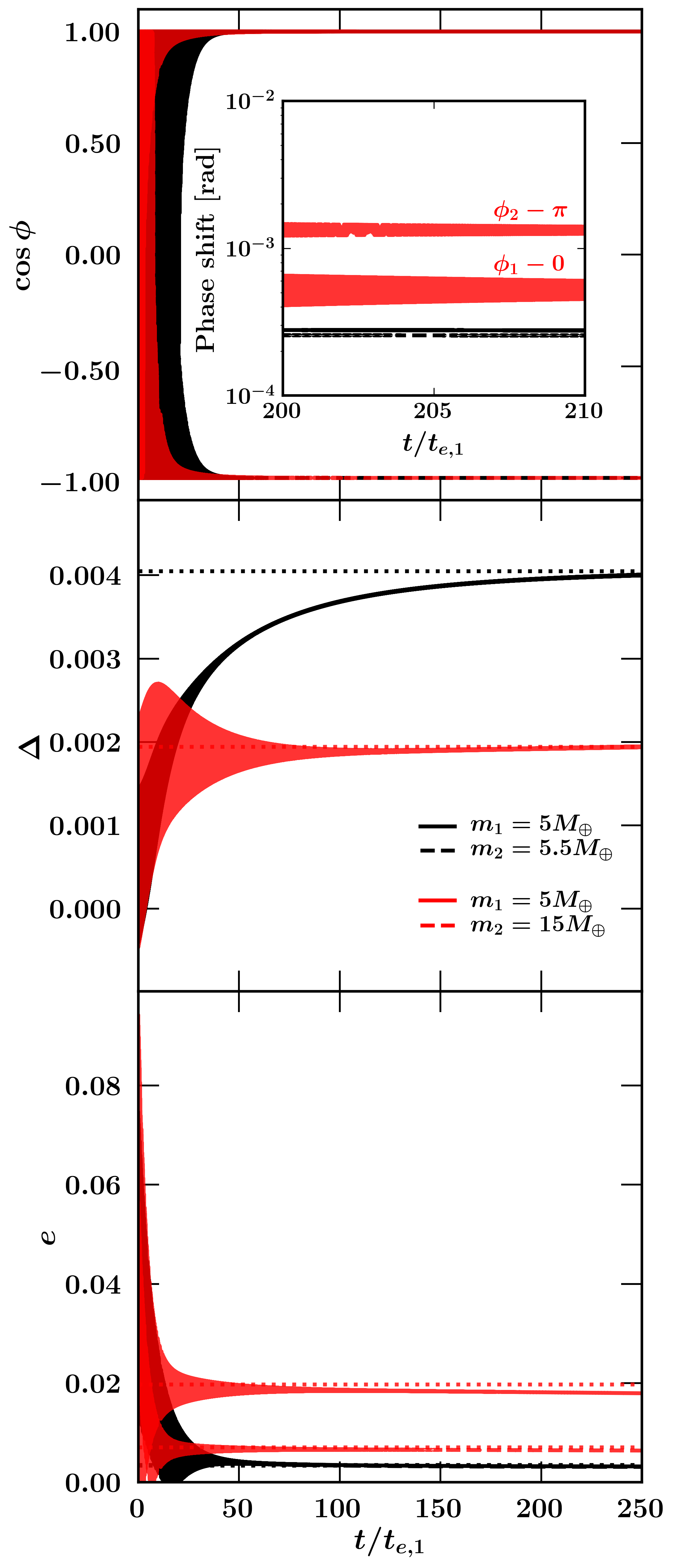}
\caption{Analogous to \autoref{fig:var_tae}, but for convergently migrating planet pairs
with non-zero masses (lifting the test particle restriction).
Two sets of mass pairings are considered 
(black vs.~red curves). Initial conditions for both sets
are as follows: 
$e_{\rm 1,initial}=e_{\rm 2,initial} = 0.05$,
$\phi_{\rm 1,initial} = 1$ rad, $\phi_{\rm 2,initial} = 2$ rad, $a_{\rm 1,initial} = 0.3$ au, and $\Delta_{\rm initial} = 0$.
We use our fiducial disc parameters 
except that we fix $\Sigma = 10$ g/cm$^2$ for
simplicity (we do not let the disc decay)
and terminate the integration when $t/\teo = 250$. 
At this end time, we analytically evaluate
equilibrium values $\Delta_{\rm eq}$,
$e_{\rm eq,1}$, and $e_{\rm eq,2}$ 
using equations (\ref{eqn:tp1})--(\ref{eqn:delta_dot}),
and plot them as horizontal dotted lines in the
middle and bottom panels.
The behaviours seen here for the unrestricted
problem are essentially the same as in
\autoref{fig:var_tae} for the restricted
problem: the bodies
lock into resonance ($\phi_1$ to $0^+$ and $\phi_2$ to $\pi^+$),
and the relative separation $\Delta$
and eccentricities $e_1$ and $e_2$ equilibrate
as expected.
}
  \label{fig:mratio}
\end{figure}

Solutions to the unrestricted problem where both planets
have non-zero mass 
and are torqued by the disc 
are qualitatively similar to those in the
restricted case. \autoref{fig:mratio} displays 
sample evolutions of two such pairs that convergently
migrate and capture into mutual resonance, with $\phi_1$
and $\phi_2$ driven to values near 0 and $\pi$, respectively. Both pairs attain equilibrium eccentricities and separations
that agree with those calculated analytically from
equations (\ref{eqn:tp1})--(\ref{eqn:delta_dot}).

\begin{figure}
\includegraphics[width=0.9\columnwidth]{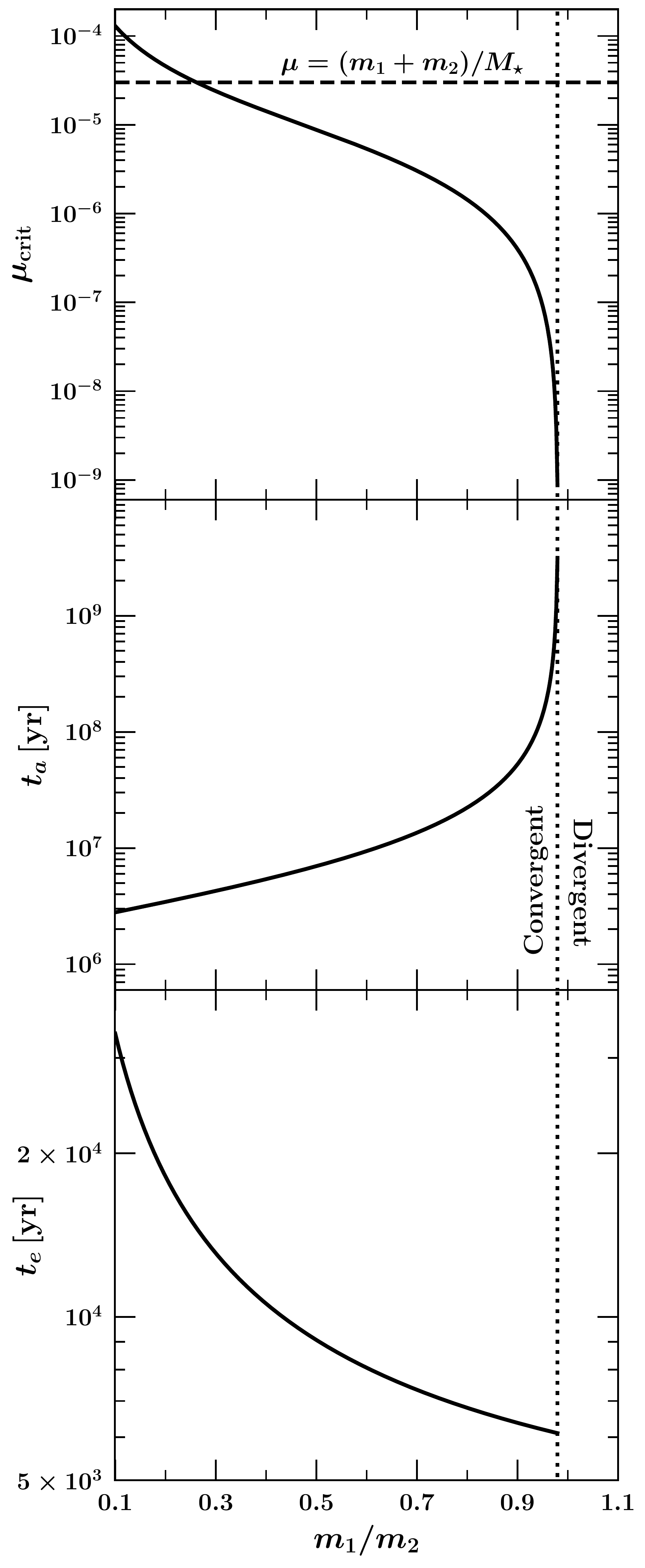}
\caption{At fixed $m_1 + m_2 = 10 \Mearth$
and fixed $\Mstar = 1 \Msun$,
the stability of the resonance 
depends on how mass is distributed between
$m_1$ and $m_2$. The top panel shows the combined planet-to-star
mass ratio $\mu_{\rm crit}$ (solid curve) above
which resonance capture is permanent and below which it is not,
computed from equation (21) of \citet{deck_batygin_2015} 
for the 3:2 resonance using our fiducial disc parameters
at $a = 0.3$ au.
As $m_1/m_2$ increases toward unity,
$\mu_{\rm crit}$ decreases, i.e., 
the threshold for stability is easier to satisfy
as we transition from
the restricted to the unrestricted three-body problem.
The critical value $\mu_{\rm crit}$ scales
as $(\te/\ta)^{3/2}$, where $\ta$ is the timescale
for relative migration ($1/\ta = 1/\tat - 1/\tao$, middle panel)
and $\te$ is a weighted average of $\teo$ and $\tet$
($1/\te = 1/\teo + (m_1/m_2)/\tet$, lower panel).
Most of the variation in $\mu_{\rm crit}$ stems from
$\ta$, which diverges as $m_1$ approaches $m_2$
(for $m_1 > 0.97 m_2$, migration is divergent and permanent
capture is not possible).
}
  \label{fig:stability}
\end{figure}

\autoref{fig:stability} explores how the stability of the resonance
changes when going from the restricted to the unrestricted problem.
We plot $\mu_{\rm crit}$, the combined planet-to-star mass ratio
above which a convergently migrating pair can stay captured in
resonance (equation 21 of \citealt{deck_batygin_2015}), vs. $m_1/m_2$
at fixed $m_1 + m_2$. The requirement for stability is easier
to satisfy ($\mu_{\rm crit}$ is smaller) in the unrestricted regime where $m_1/m_2$ is near unity.
The variation in $\mu_{\rm crit}$
arises mostly from its dependence on 
$\ta = \tao\tat/(\tao - \tat)$, the relative migration timescale.
As the planets become more comparable in mass, their individual
timescales for migration $\tao$ and $\tat$ approach each
other, $\ta$ increases, and by extension $\mu_{\rm crit}
\propto 1/\ta^{3/2}$ decreases. This effect helps to stabilize
the population of sub-Neptunes we mock up in \autoref{sec:mc}.

\begin{figure*}
\includegraphics[width=0.99\textwidth]{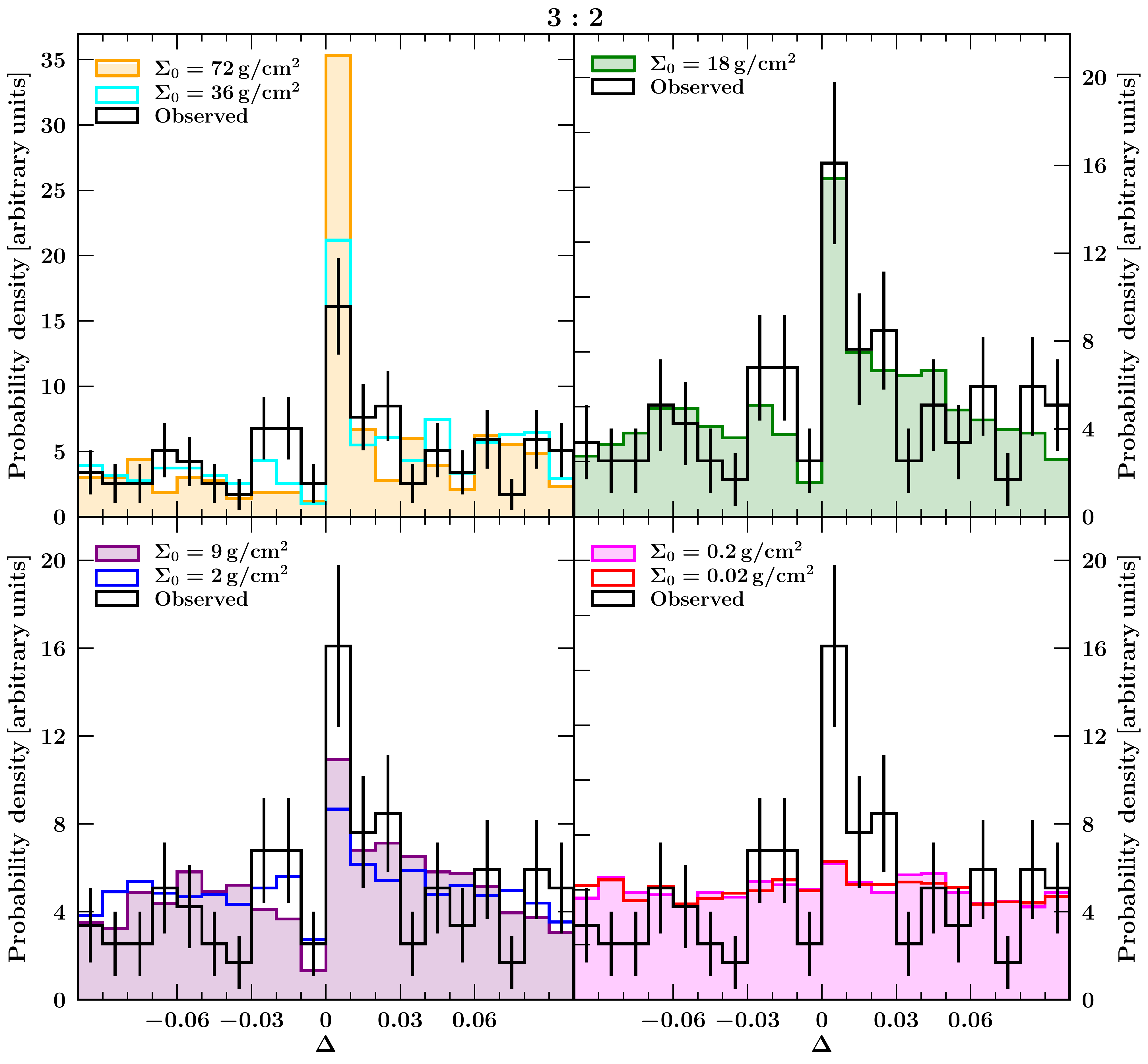}
\caption{Final distributions of $\Delta$ from our Monte Carlo population synthesis model of the 3:2 resonance (open and filled coloured 
histograms) compared against the observed distribution (black open histogram). The 3:2 observations are culled of pairs
with inner planet periods $P_1 < 5$ days to avoid potential
contamination from stellar tidal interactions; see Figures \ref{fig:delta_psplit} and \ref{fig:cdf}. Of the models shown, the one corresponding
to an initial disc gas surface density of $\Sigo = 18$ g/cm$^2$ (upper right in green) reproduces
the observed peak-trough
feature best. This surface density, which characterizes our
model disc from 0.1 to 1 au, is 3--4 orders
of magnitude lower than corresponding surface densities
in minimum-mass, solar-composition
reconstructions of protoplanetary discs derived
from {\it Kepler} data (e.g., \citealt{chiang_laughlin_2013}).
In our model, the peak mostly comprises planet pairs
that convergently migrate from $\Delta_{\rm initial} > 0.01$
to an equilibrium separation
$\Delta_{\rm eq} \simeq 0.001$--0.01.
Values for $\Sigo \gtrsim 18$ g/cm$^2$ lead to more migration
and overpredict the number of systems captured into the peak;
conversely, $\Sigo \lesssim 18$ g/cm$^2$
underpredicts the peak.}
  \label{fig:mc_master}
\end{figure*}

\section{Population synthesis}
\label{sec:mc}
Having reconnoitered the outcomes of two planets
undergoing semimajor axis and eccentricity changes
near resonance, we now construct a population synthesis model 
designed to reproduce
the observed distribution of $\Delta$'s near the 3:2 resonance. 
The goal is to identify the disc conditions---in particular
the disc gas surface density around the time sub-Neptunes
attain their final masses---that are compatible with the
$\Delta$-distribution, and thereby assess the degree to which
planets migrated. As our model is 
simplistic, 
the most we can hope for is that our inferences
will be accurate enough to point us in the right direction
when thinking about sub-Neptune formation---whether to
a gas-rich disc where such planets typically 
migrate large distances,
or to a gas-poor one where they spawn more-or-less
in situ.

\subsection{Monte Carlo method}\label{ssec:popsynth}

Our calculation of the dynamical evolution of a pair of
sub-Neptunes begins just after they form, i.e., just after their solid cores, which dominate their masses, coagulate. At this time ($t=0$), 
the disc gas surface density $\Sigma$ everywhere equals 
$\Sigo$ (assuming $\beta_\Sigma = 0$);
thereafter, $\Sigma$ decays exponentially (equation \ref{eqn:sigo}). 
We consider values for $\Sigo$ between 
0.01 and 100 g/cm$^2$, a range
that we will see brackets the best fit to the
observations.
For every value of $\Sigo$ chosen, we integrate the equations of
motion (\ref{eqn:n1dot})--(\ref{eqn:phi2dot}) 
until $\Sigma$ has decreased by 
two orders of magnitude relative to its initial value.
Thus, for example, a model 
with $\Sigo = 10$ g/cm$^2$ is integrated from
$\Sigma = 10$ g/cm$^2$ to $\Sigma = 0.1$ g/cm$^2$.
We have verified that integrating further changes 
our results negligibly. 

For every $\Sigo$, we construct $N=2000$ planetary systems
with properties and initial conditions chosen randomly as follows.
Host stellar masses are drawn uniformly from 0.5 to 2 $\Msun$, 
approximately 
matching the range of masses reported in the NASA Exoplanet Archive.
For every star we lay down two planets whose masses are each 
drawn randomly from a uniform distribution between 5 and 15 $\Mearth$ (cf.~\citealt{lithwick_etal_2012, weiss_marcy_2014, hadden_lithwick_2014, hadden_lithwick_2017, wu_2019}). 
The inner planet is initialized with a
semimajor axis $a_{\rm 1, initial}$ chosen randomly from a distribution that is
uniform in $\log a_{\rm 1, initial}$ between 0.1 and 1 au (corresponding to
orbital periods of $\sim$10 to $\sim$400 days);
such a distribution is similar to that observed \citep[e.g.,][]{fressin_etal_2013, dressing_charbonneau_2015}.
We set the outer planet's semi-major axis 
such that the pair are near commensurability: 
we draw $\Delta_{\rm initial}$ from a flat distribution
between -0.1 and 0.1, a range that encompasses the
observed period ratio asymmetry (e.g., \autoref{fig:delta_psplit}). Together, $\Delta_{\rm initial}$ and $a_{\rm 1,initial}$ specify
the initial location of the outer planet:
\begin{equation}
    a_{\rm 2, initial} = \left[\left(\Delta_{\rm initial}+1\right)\frac{q+1}{q}\right]^{2/3}a_{\rm 1,initial} 
\end{equation}
with $q=2$ for the 3:2 resonance. 
Initial eccentricities $e_{\rm 1, initial}$ and $e_{\rm 2, initial}$
are each drawn randomly from a uniform distribution between
0 and 0.1, and initial resonant arguments $\phi_{\rm 1,initial}$ and $\phi_{\rm 2, initial}$
are each drawn randomly from a uniform distribution between
0 and $2\pi$.
In general the planets do not begin in resonance.

The semimajor axis and eccentricity driving terms from
the background disc are given
by equations (\ref{eqn:ta})--(\ref{eqn:te}). They
and our input parameters---in particular
our nominal choice for $\beta_\Sigma = 0$---are such that 
while all planets migrate inward, 
about 50\% of planet pairs convergently
migrate ($\tat < \tao$),
with the remaining fraction migrating divergently.
Only a convergent pair can capture into resonance
and attain an equilibrium separation $\Delta_{\rm eq} > 0$
(see equation \ref{eqn:delta_dot}, and Sections \ref{sec:restricted} and \ref{sec:unrestricted}).

For every $\Sigo$, we compare the $N= 2000$ final values of
$\Delta$ ($=\Delta_{\rm final}$) against the observed $\Delta$-distribution.
Should a planet migrate to the inner edge of the disc, which we take to lie at 
$P = 3$ days \citep[e.g.,][]{elee_chiang_2017},
we shut off the disc torque acting on the planet, 
setting its $\ta$ and 
$\te$ to infinity. An inner planet so stopped at the edge may be pushed 
further inward by resonant interaction with the outer planet; 
this process stops once the outer planet also hits the disc inner edge,
at which point we terminate the integration. 
We also halt an integration if both planets convergently migrate
such that their semimajor axes coincide ($\Delta$ near -1/3).
For the models that best fit the observations,
none of these eventualities is significant.

\subsection{Comparison to observations}
\label{ssec:mc_results}

\begin{figure}
\includegraphics[width=0.95\columnwidth]{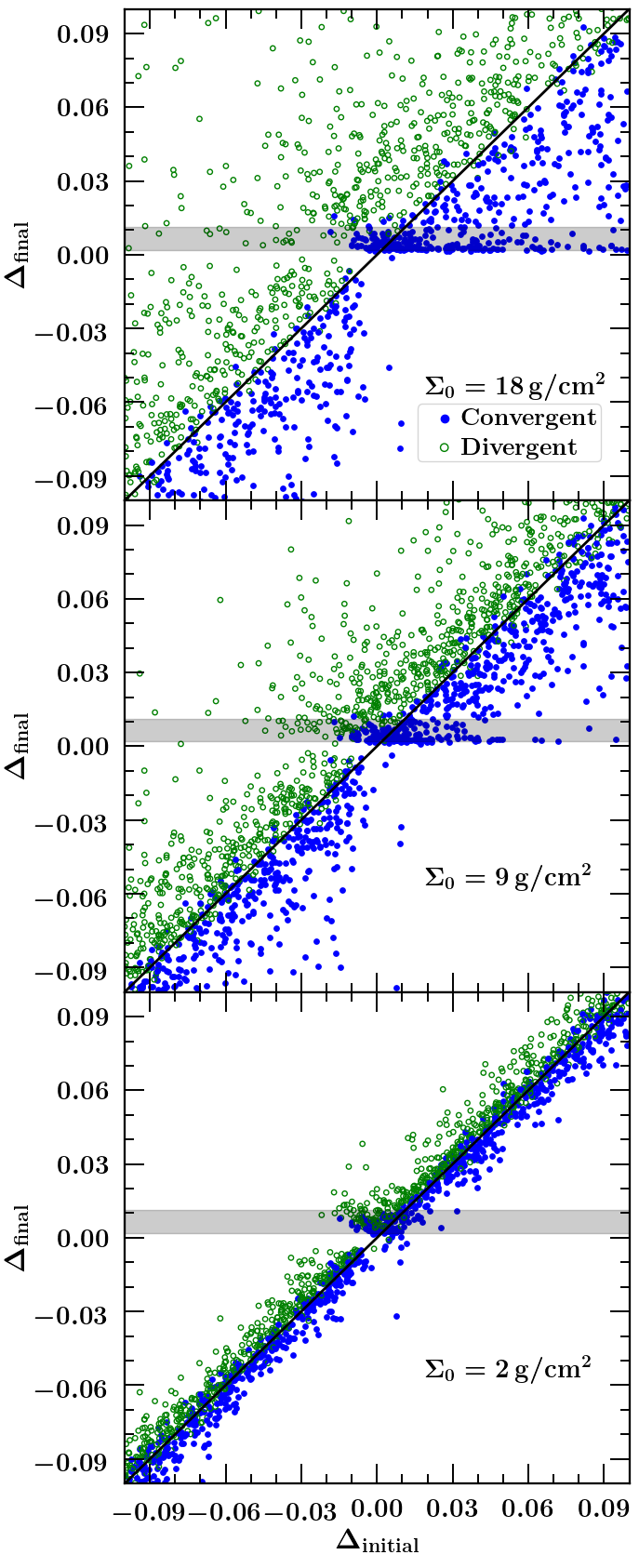}
\vspace{-0.3cm}
\caption{How initial separations $\Delta_{\rm initial}$
map to final separations $\Delta_{\rm final}$, for
three population synthesis models with varying 
initial disc surface densities $\Sigo$.
Away from resonance, convergently migrating pairs (blue points)
move down, below the slope 1 line, to 
$\Delta_{\rm final} < \Delta_{\rm initial}$; 
divergently migrating pairs (green points) move up 
to $\Delta_{\rm final} > \Delta_{\rm initial}$. The greater
is $\Sigo$ (the more massive the disc),
the more systems move in $\Delta$. Convergent pairs that start at $\Delta_{\rm initial} > 0.01$ and migrate to $0.001 \lesssim \Delta_{\rm final} \lesssim 0.01$ (grey shaded band computed
from equation \ref{eqn:delta_dot} using our Monte Carlo
inputs) stay there, trapped near stable resonant
fixed points. The peak
in the $\Delta$-histogram (\autoref{fig:mc_master})
is largely made up of these systems.
Convergent systems can also start from
$-0.01 \lesssim \Delta_{\rm initial} < 0$
and become resonantly trapped in the grey band at $\Delta_{\rm final} > 0$
by eccentricity damping. For more details,
including a discussion of divergent systems, see
\autoref{ssec:mc_results}.}
  \label{fig:bird}
\end{figure}

\begin{figure}
\includegraphics[width=\columnwidth]{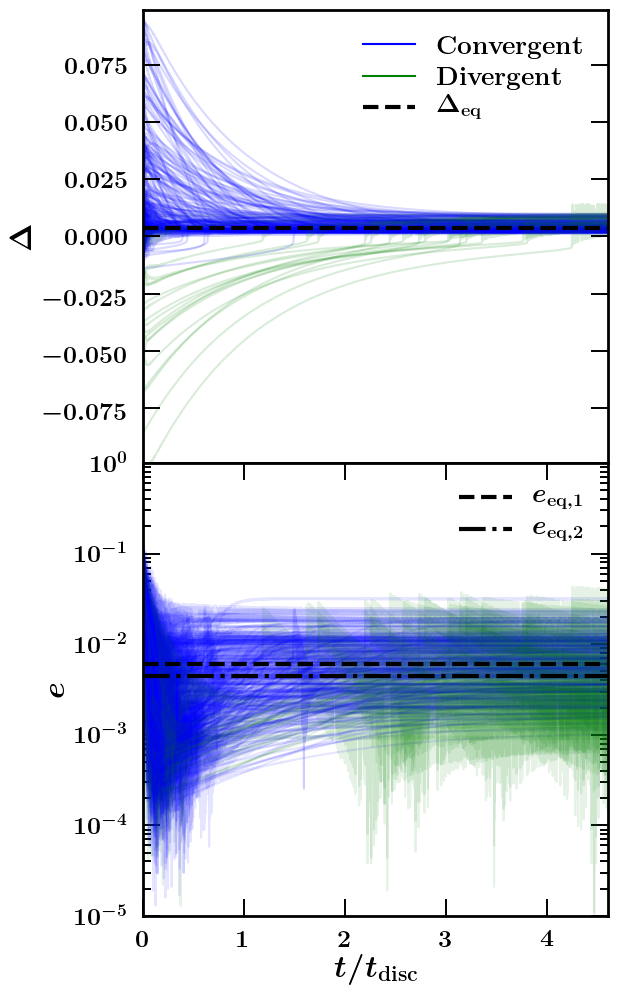}
\vspace{-0.5cm}
\caption{How systems in the peak of the $\Delta$-distribution
arrived there. Only planet pairs in our $\Sigo = 18$ g/cm$^2$
model with $0 < \Delta_{\rm final} < 0.01$ (i.e., the peak;
\autoref{fig:mc_master}) are plotted here. 
Most systems land in the peak by migrating convergently 
from $\Delta_{\rm initial} > 0.01$ and settling into resonant
equilibria (blue curves); a few migrate divergently 
from $\Delta_{\rm initial} < 0$ and are left in the peak
when the disc disperses (green curves). 
Horizontal dashed line in the top panel gives the median $\Delta_{\rm eq}$ computed from \autoref{eqn:delta_dot} 
using our Monte Carlo inputs for convergent pairs only.
Horizontal lines in the bottom panel give
median equilibrium eccentricities computed similarly
from equations (\ref{eqn:tp1})--(\ref{eqn:tp2}).}
  \label{fig:brown}
\end{figure}

\autoref{fig:mc_master} shows the final
$\Delta$-distribution as a function of the initial gas surface density $\Sigo$. For every $\Sigo$ tested,
we see an excess number of systems with
$0 < \Delta_{\rm final} < 0.01$. Most of this excess population
comprises convergently
migrating planet pairs that capture into resonance
and equilibrate in $\Delta$
(as described in Sections 
\ref{sec:restricted} and \ref{sec:unrestricted}).
This equilibration is illustrated 
in Figures \ref{fig:bird} and \ref{fig:brown},
where we see the convergently migrating systems,
colored in blue, converge on
$\Delta_{\rm final} = \Delta_{\rm eq} \approx 0.005$.
As a check on our numerics, 
we compute independently the value 
of $\Delta_{\rm eq}$ using equation
(\ref{eqn:delta_dot}),
finding $\Delta_{\rm eq} \simeq 0.001$--0.01 (10th--90th
percentile range) for our Monte Carlo inputs; 
this range is plotted as a horizontal shaded
bar in Figures \ref{fig:bird} and \ref{fig:brown}
and agrees well with the data.

The excess population
at $\Delta_{\rm eq}$---the ``peak''
in the $\Delta$-histogram in \autoref{fig:mc_master}---increases with the number of systems that 
migrate convergently (with $\dot{\Delta} < 0$) from
large $\Delta_{\rm initial} > 0.01$ to $\Delta_{\rm eq}$ over the disc lifetime.
The more massive the disc, the faster the relative migration
and the wider the range of $\Delta_{\rm initial} > 0.01$ 
that the peak draws from.
The height of the peak relative to the background
is reproduced approximately by $\Sigo = 18$ g/cm$^2$
(\autoref{fig:mc_master}, top right panel).
Less massive discs (bottom
panels) produce too small a peak,
and more massive discs (top left panel) too large.

Accompanying the peak is a ``trough''---a deficit
of systems with $-0.01 < \Delta_{\rm final} < 0$.
The observed trough relative to the background continuum
appears reproduced by $\Sigo = 2$--18 g/cm$^2$ (\autoref{fig:mc_master}).
The trough is created by both convergent and divergent pairs
with $-0.01 < \Delta_{\rm initial} < 0$ moving
to $\Delta_{\rm final} > 0$ (\autoref{fig:bird}).
The crossing to positive $\Delta$
is effected by disc-driven eccentricity damping
and not by disc-driven migration, as the same transport
occurs when we turn off the latter.
Moreover, numerical experiments show that 
the width of the trough---the range of negative
$\Delta_{\rm initial}$ over which systems are 
transported---increases linearly with
planet mass, presumably reflecting how the
eccentricity damping rate scales linearly with planet mass. 
The convergent pairs that are transported become 
permanently captured into resonance and equilibrate at
$\Delta_{\rm eq} > 0$ (see the data colored
blue in Figures \ref{fig:bird} and \ref{fig:brown}). 
Divergent systems that are transported
(Figure \ref{fig:bird}, green points) 
do not permanently lock and equilibrate,
but continue toward increasing $\Delta$.
The small fraction of divergent systems
that contribute to the peak 
(10\% for $\Sigo = 18$ g/cm$^2$; \autoref{fig:brown},
green lines) represent pairs that were merely ``passing 
through'' the peak in $\Delta$-space when they ``froze''
in place with the dispersal of the disc. 

Our constraints on $\Sigo$, which are based on reproducing the observed $\Delta$-distribution, 
depend on the disc dispersal time $t_{\rm disc}$, as
the amount by which 
systems are transported in $\Delta$-space scales as the product $\Sigo \times t_{\rm disc}$. 
Thus our best-fit $\Sigo = 18$ g/cm$^2$, which pairs
with our nominal $t_{\rm disc} = 10^5$ yr, is degenerate
with $\Sigo = 1.8$ g/cm$^2$ and $t_{\rm disc} = 10^6$ yr.

\begin{figure}
\includegraphics[width=\columnwidth]{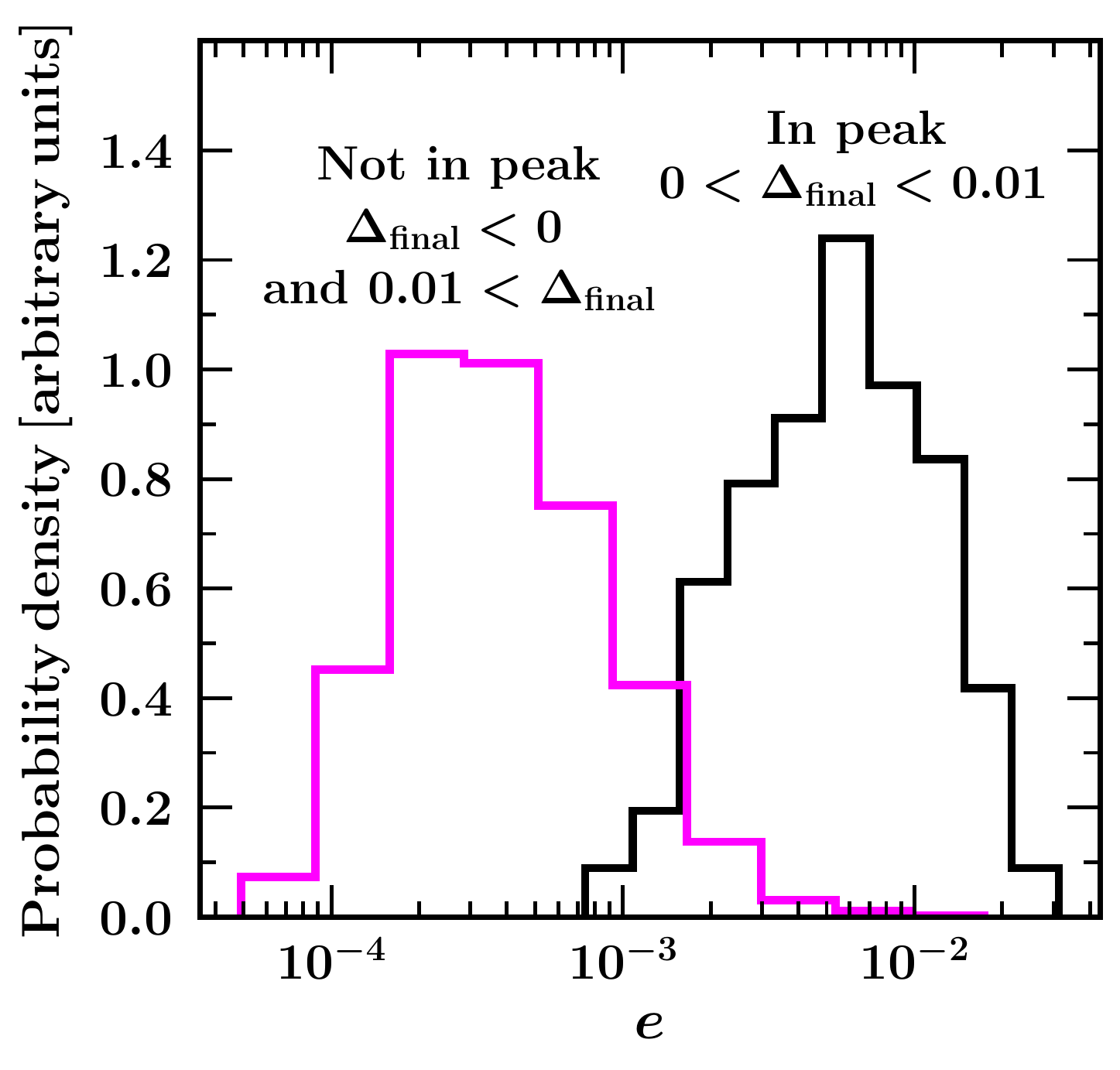}
\vspace{-0.5cm}
\caption{Final eccentricities of planets within
the peak of the $\Delta$-distribution (black histogram) and outside the peak (magenta histogram),
for our best-fitting $\Sigo = 18$ g/cm$^2$ population synthesis model. Within the peak, eccentricities
amplify from resonant migration and damp
from first-order co-orbital Lindblad torques, reaching equilibrium values of $10^{-3}$--$10^{-2}$. Outside the peak, there is no resonant amplification
of eccentricity, only disc damping.}
  \label{fig:ehist}
\end{figure}

Convergently migrating pairs comprising the peak
equilibrate not only in $\Delta$ but also in $e$,
as eccentricity pumping by resonant migration balances
eccentricity damping by the disc. \autoref{fig:brown}
shows that eccentricities of planet pairs that convergently
migrate into resonance starting from  $\Delta_{\rm initial} > 0.01$ equilibrate to $e_{\rm eq} \sim 10^{-3}$--$10^{-2}$
(lower panel, blue curves). 

The handful of systems that find their way into the peak from
$\Delta_{\rm initial} < -0.01$ exhibit similar final eccentricities,
but via a different path:
just after the planets cross into resonance on diverging orbits
(effected either by eccentricity damping or divergent migration),
their eccentricities jump to values $\gtrsim 0.01$
\citep{dermott_etal_1988}, 
after which they damp back down by residual disc torques.

More generally, systems in the peak
have eccentricity histories that differ from systems
outside the peak, as the former have been influenced
by the resonance whereas the latter have not. \autoref{fig:ehist}
shows that, for our assumed Monte Carlo inputs (and neglecting
post-formation gravitational interactions lasting
Gyrs), eccentricities of non-peak systems
are systematically lower than for systems in the peak.

\begin{figure*}
\includegraphics[width=\textwidth]{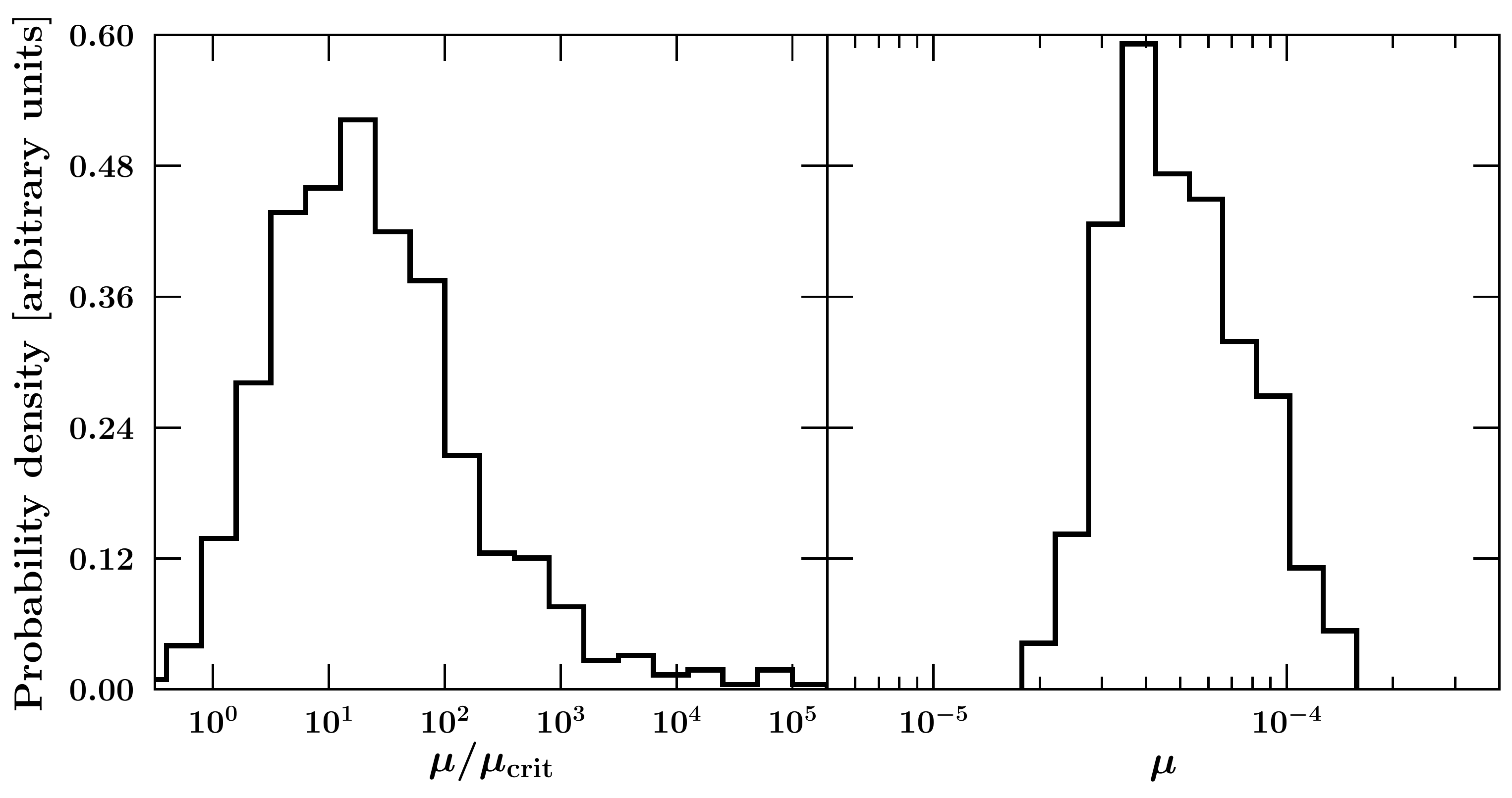}
\vspace{-0.5cm}
\caption{Distributions of $\mu/\mu_{\rm crit}$ (left panel) and $\mu$
  (right panel), where $\mu = (m_1 + m_2)/\Mstar$ and $\mu_{\rm crit}$
  is the critical value below which the 3:2 resonance is unstable, for all
  convergently migrating pairs in our mock population. Our modeled
  sub-Neptunes, having masses between 5 and $15 M_\oplus$, and
  orbiting stars between 0.5 and $2 M_\odot$, are stable.}
  \label{fig:mu_hist}
\end{figure*}

In our Monte Carlo calculations, convergently migrating pairs that
capture into resonance stay in resonance. Escape
after capture does not occur---in \autoref{fig:bird}, systems
that convergently migrate into the shaded horizontal bar
denoting $\Delta_{\rm eq}$ stay there.
We checked this result
by comparing our combined planet-to-star mass ratios,
$\mu \equiv (m_1+m_2)/M_\star$, to the critical value
$\mu_{\rm crit}$ below which a system escapes resonance
(equation 21 of \citealt{deck_batygin_2015}).
The left panel of \autoref{fig:mu_hist} shows the distribution of $\mu/\mu_{\rm crit}$
for all convergently migrating pairs in our mock planet
population. The distribution peaks at $\mu/\mu_{\rm crit}
\sim 10$ and extends to values even larger---a consequence of comparable masses $m_1$ and $m_2$ leading to a longer relative
migration timescale $\ta$ and thus a smaller
$\mu_{\rm crit} \propto 1/\ta^{3/2}$ (\autoref{sec:unrestricted})---demonstrating that
practically all our simulated resonances
are stable. 
Our finding contrasts with an earlier
suggestion by \citet[][GS]{goldreich_schlichting_2014} that
sub-Neptunes generically escape resonance (their figs.~10 and 11).
The difference stems in part from our respective disc models: the GS model drops order-unity factors and
fixes $\ta / \te \equiv 3 n / (2\dot{n} \te) = 3 (h/a)^{-2}/2 = 150$,
whereas ours retains order-unity factors and $a$-dependencies
(equations \ref{eqn:ta}--\ref{eqn:sigo}) to find that $\ta/\te$ for an individual planet varies from 200 to 750 across the disc. 
Consequently, as $\mu_{\rm crit} \propto (\te/\ta)^{3/2}$,
our values for $\mu_{\rm crit}$ are systematically lower than theirs by
factors of 1.5--11. Furthermore,  
our combined planet-to-star mass ratios $\mu$, shown in the right panel of \autoref{fig:mu_hist}, are greater than the values adopted by GS by an order of magnitude
($\mu = 10^{-5}$--$10^{-4}$
vs. $10^{-6}$--$10^{-5}$). Our planet masses are modeled
after radial velocity measurements \citep[e.g.,][]{weiss_marcy_2014},
transit timing variations 
\citep{lithwick_etal_2012,hadden_lithwick_2014,hadden_lithwick_2017},
and evolutionary models for planet radius \citep[][]{wu_2019},
whereas the GS values derive from planet radius measurements
from 2013 and an assumed universal bulk density of 2 g/cm$^3$.

\begin{figure*}
\includegraphics[width=0.99\textwidth]{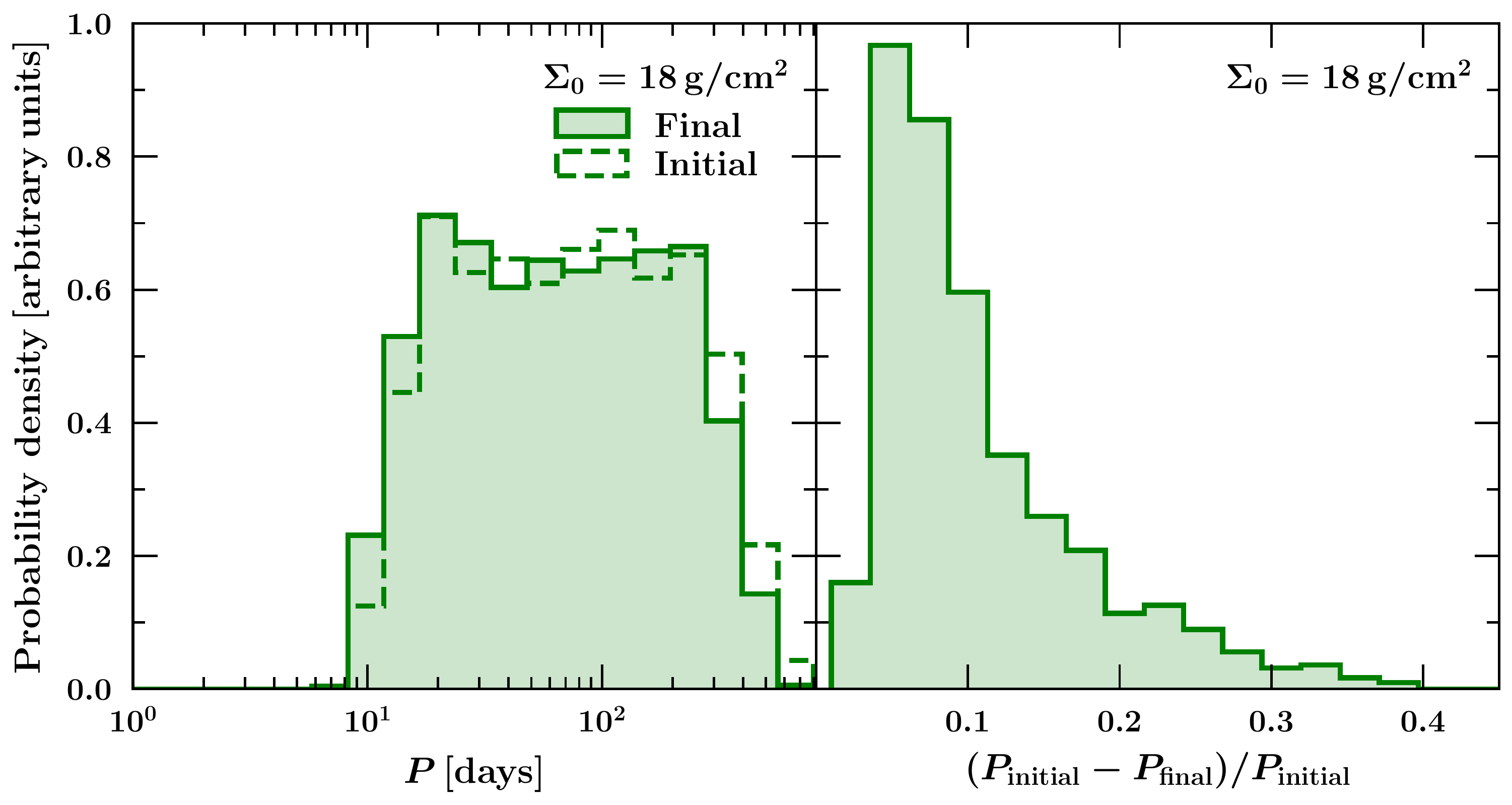}
\caption{Final vs.~initial period distributions of our mock planet
  population (left panel), and the fractional changes to the periods
  (right panel), for our best-fit $\Sigo = 18$ g/cm$^2$
  model. Migration is modest; the median change in orbital period is
  $\sim$10\%.}
\label{fig:P_master}
\end{figure*}

While the relative separation $\Delta$ can
increase, decrease, or equilibrate, all planets
migrate inward (by construction from our
adoption of Type I migration).
\autoref{fig:P_master} compares the final distribution
of individual planet periods to the initial
distribution for $\Sigo = 18$ g/cm$^2$,
the value that gives 
an encouraging match to the observed $\Delta$-distribution
(\autoref{fig:mc_master}).
There is hardly any migration: the median change
in orbital period is 10\% (\autoref{fig:P_master},
right panel). Insofar as these changes are small,
sub-Neptunes can be said to have completed
their mass assembly more-or-less in situ.

\begin{figure*}
\includegraphics[width=\textwidth]{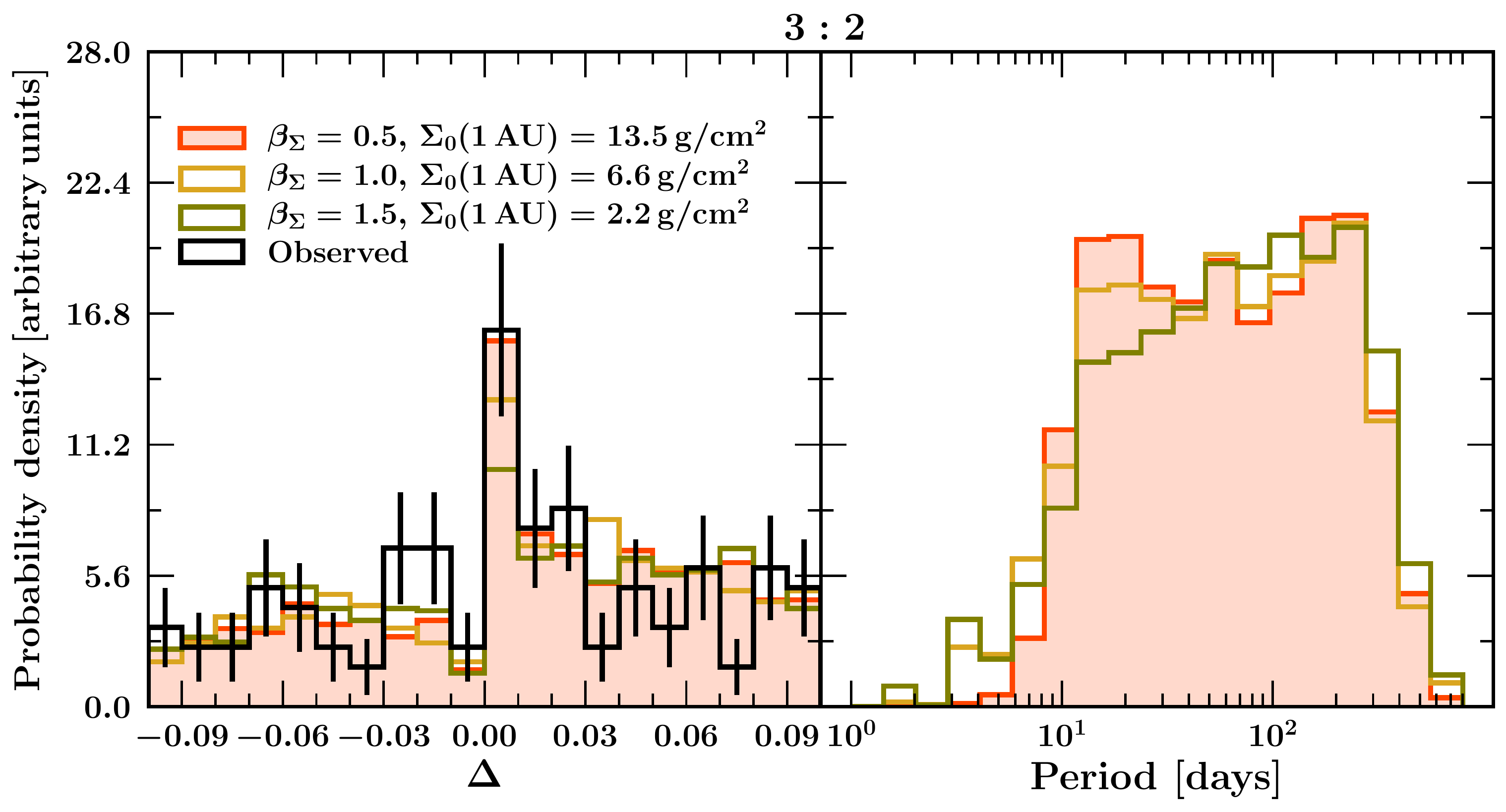}
\vspace{-0.5cm}
\caption{Attempts to reproduce the peak-trough asymmetry using
  alternate disc surface density profiles. Steeper profiles (larger
  $\beta_{\Sigma} = -d\log \Sigma/d\log a$) lead to inner planets
migrating faster and fewer convergently migrating pairs, reducing
the number of systems that populate the peak (left panel).
With steeper profiles, more planets migrate toward the disc inner edge, producing
pile-ups in the occurrence rate at short period (right panel)
that are not observed \citep{elee_chiang_2017}.}
  \label{fig:beta}
\end{figure*}

Finally, in \autoref{fig:beta}, we relax
our standard assumption of a flat gas surface density
profile and experiment with
$\beta_\Sigma \equiv - d\log \Sigma / d\log a \neq 0$,
with $\Sigo$ now denoting the surface density at
$a = 1$ au only (equation \ref{eqn:sigo}).
As $\beta_\Sigma$ increases and the gas
density toward the star rises, inner planets
migrate faster and fewer planet pairs migrate convergently. 
Consequently the simulated peak
in the $\Delta$-distribution weakens (left panel).
Higher $\beta_\Sigma$ also leads to more inner planets
migrating to orbital periods $P < 10$ days (right panel),
toward the innermost disc edge at $P = 3$ days.
Migrating toward the disc edge is not desirable
as it leads to pile-ups in the occurrence rate
at short $P$ that are not observed
\citep{elee_chiang_2017}.
We have tried to mitigate against pile-ups 
by reducing $\Sigo$ as $\beta_\Sigma$ increases 
in \autoref{fig:beta}; however,
the surface densities
employed in this figure still resemble
those of the disfavored migration models
of \citet{elee_chiang_2017}, which betray 
pile-ups that do not compare
well with the observed sub-Neptune period distribution
(see their figs.~4 and 5).
\rev{In this regard the data appear to prefer
surface densities that either stay
flat or increase away from the star---these yield
more convergent pairs and healthier
peaks (\autoref{fig:mc_master}),
while keeping migration-induced pile-ups
at bay (\autoref{fig:P_master}).
Other studies have independently come to this same
conclusion, as we discuss in \autoref{sec:summary_discussion}.}

\section{Summary and discussion}
\label{sec:summary_discussion}

\subsection{Gas-Rich vs.~Gas-Poor Discs, Migration vs.~In-Situ Formation}
\label{sec:sum1}

Most pairs of sub-Neptunes and super-Earths
are not in mean-motion
resonance (\autoref{fig:period_ratio_distributions};
\citealt{lissauer_etal_2011, fabrycky_etal_2014}).
At face value, this observation
suggests that disc-driven migration plays only a
limited role in the formation of such planets, as
wholesale changes to orbital periods would be expected
to capture a large fraction of bodies into resonance. \citet[][]{goldreich_schlichting_2014}
warned against
this conclusion, pointing out that
sufficiently low-mass planets escape from resonance 
when their eccentricities are damped by interactions with their natal discs.
While our calculations confirm the existence
of a planetary system mass below which
resonances are overstable,
we find that most
sub-Neptune pairs sit above this threshold, \rev{and their resonances are therefore
stable against eccentricity damping.} Current inferences of sub-Neptune masses
from radial velocity measurements
\citep[e.g.,][]{weiss_marcy_2014},
transit timing variations \citep{lithwick_etal_2012,hadden_lithwick_2014,hadden_lithwick_2017},
and theoretical radius-mass relations incorporating
photoevaporative mass loss \citep[][]{wu_2019}
indicate planet-to-star mass ratios at least a factor of
$\sim$10 higher than the values used by \citet{goldreich_schlichting_2014}; compare our \autoref{fig:mu_hist} to their fig.~10.

While most planet pairs are not in resonance,
some are. There are excess numbers of systems
just wide of resonance, with $0 < \Delta \lesssim 0.02$, 
where $\Delta$ is the fractional deviation
of a pair's period ratio away from 3/2 or 2/1 (equation \ref{eqn:delta_def}).
Accompanying these excesses are deficits
in the planet population just short
of resonance, with $-0.02 \lesssim \Delta < 0$.
We have sought to reproduce 
this ``peak-trough'' feature in the $\Delta$-histogram by modeling 
the dynamical evolution of planets
within their natal discs. 
In our model, the peak mostly comprises
pairs of planets each of which convergently migrated
from $\Delta_{\rm initial} > 0.01$, captured
into resonance, and attained an equilibrium
wherein migration-induced, resonant amplification
of eccentricities balances disc damping
of eccentricities. This eccentricity equilibrium
corresponds, in resonance,
to an equilibrium in relative semi-major axes, i.e.,
an equilibrium in $\Delta$ \citep[e.g.,][]{terquem_papaloizou_2019}.
For our model parameters, 
$\Delta_{\rm eq} = 0.001$--0.01, which matches well
the position of the peak for the 3:2 resonance,
at least for periods longer than $\sim$5 days
(more on the 2:1 resonance in \autoref{sec:sum2} below). The trough corresponds to 
systems that begin at $-0.01 < \Delta_{\rm initial} < 0$ and are transported into resonance at $\Delta > 0$ by disc eccentricity damping.
The parameters responsible for this quantitative
agreement include disc
aspect ratios of $h/a \simeq 0.02$--0.04, appropriate
for the $a = 0.1$--1 au orbital distances of
{\it Kepler} planets, and planet-pair-to-star mass
ratios of $\mu \sim 10^{-5}$--$10^{-4}$.

The more massive the disc 
and the longer it persists, 
the farther away a pair can be from resonance
(i.e., the larger $\Delta_{\rm initial}$ can be) 
and still be brought into resonant contact by migration---more
migration brings more systems into the
peak. The observed peak and trough for the 3:2 resonance are approximately
reproduced for a disc e-folding time of $t_{\rm disc} = 10^5$ yr
and a gas surface density of $\Sigo \sim 20$ g/cm$^2$ at orbital
distances $a = 0.1$--1 au (\autoref{fig:mc_master}). 
An equivalent fit is obtained for
$t_{\rm disc} = 10^6$ yr and $\Sigo \sim 2$ g/cm$^2$.

\rev{Gas surface density profiles that are flat if not
actually rising with distance from the host star are preferred as
they lead to a greater proportion of planet pairs
that migrate convergently and efficiently populate the peak.}
Gas profiles that fall steeply away from the star 
(like that of the minimum-mass nebula,
$\Sigma \propto a^{-3/2}$) are disfavored because
they lead to a smaller fraction of convergent
pairs; faster migration rates, i.e., 
higher overall surface densities, 
are then needed to reproduce the peak, but
these lead to pile-ups of sub-Neptunes near the disc
inner edge that violate observed occurrence rate profiles \citep{elee_chiang_2017}.

Since we do not follow the growth of planet masses,
but merely mock up their present-day values, 
our model gas surface densities should be interpreted as
characterizing the disc around the time
planets complete their assembly, during or just
after their last mass doubling. 
The gas densities we have inferred from fitting the peak-trough feature are low: 2--20 g/cm$^2$ is 3--5 orders of
magnitude lower than the gas content of a minimum-mass, 
solar composition nebula at the relevant orbital distances. 
Our preferred model disc appears incompatible with
gas-rich scenarios (e.g., pebble accretion) for
super-Earth and sub-Neptune formation.
Our results point instead to gas-poor formation scenarios, in particular giant impacts 
\citep[e.g.,][]{dawson_etal_2015, elee_2019, macdonald_etal_2020}.
Our fitted nebular densities are quantitatively consistent 
with the formation of super-Earth/sub-Neptune cores by 
giant impacts; gas densities 
are sufficiently low, and by extension disc eccentricity
damping is sufficiently weak, that proto-cores, each a few Earth
masses and spaced several Hill radii apart, can gravitationally stir one another onto crossing orbits and merge into full-fledged super-Earths (see, e.g., the $k=7$ curve in fig.~5 of \citealt{elee_chiang_2016}). Although we need planets
to migrate 
to produce the peak,
they should not migrate by much, \rev{as otherwise the magnitude of the peak would be overestimated}:
in our preferred model, the median change
in the orbital period of an individual planet
is only 10\% (\autoref{fig:P_master}). 
Accordingly, we would describe the endgame of sub-Neptune
formation as occurring largely in situ.
This same conclusion is reached by
\citet{terquem_papaloizou_2019}. 
Furthermore, \citet{macdonald_dawson_2018}
show that the 
handful of multi-planet resonant chains like
Kepler-223 \citep{mills_etal_2016}
and Trappist-1 \citep{gillon_etal_2017}
do not necessarily implicate wholesale
migration, but can also be established 
by more modest ``short-scale'' migration, as advocated here.

\rev{Encouragingly, our preferred low-mass disc with a
flat-to-rising surface density profile
resembles disc models computed by \citet{suzuki_etal_2016}
using strongly magnetized winds to drive accretion (see
their fig.~1, the red or green curves).
\citet{ogihara_etal_2018} have shown
that sub-Neptunes within such discs undergo their last
mass doubling/giant impact around $10^6$ yr (their
fig.~3c), at which time the surface density at 0.1--1 au 
is about $20$ g/cm$^2$ (their fig.~1a). Their calculated final period
distribution of sub-Neptunes matches that observed 
(their fig.~14), with no pile-ups from excessive migration.
The largely in-situ formation history
described by these authors agrees with ours.}

\subsection{Tides and the 2:1 and 3:2 Resonances}\label{sec:sum2}

\begin{figure}
\includegraphics[width=\columnwidth]{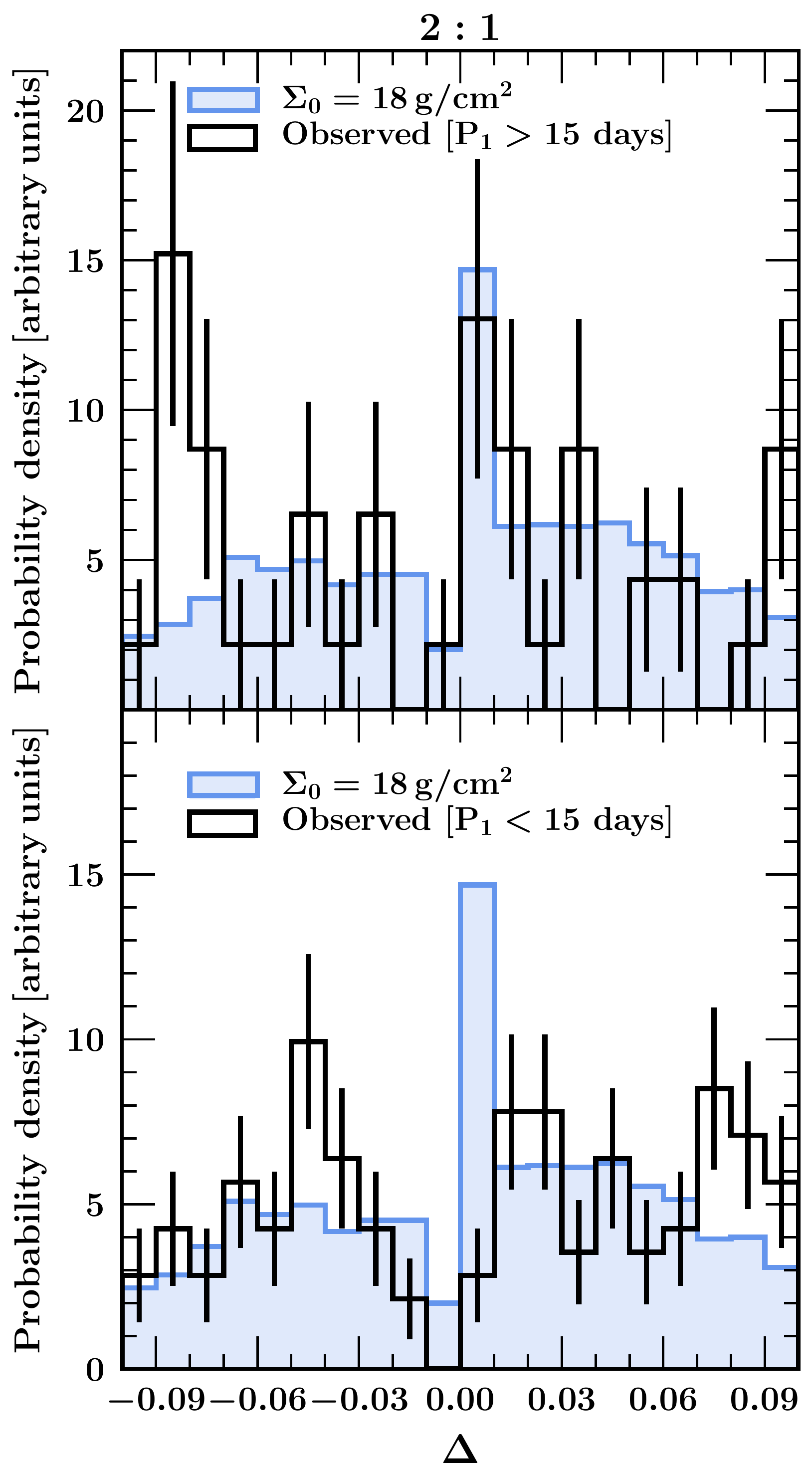}
\vspace{-0.5cm}
\caption{The $\Delta$-distribution for the 2:1 resonance,
observed (black) and modeled (blue). 
The observational data are split according to the period
of the inner planet $P_1$. 
At $P_1 > 15$ days (upper panel), 
systems are presumably least impacted by stellar tidal
interactions, and the peak-trough asymmetry
can be reproduced by our disc-only, no-tide model.
Disc parameters are the same as those fitted for the 3:2
($\Sigo = 18$ g/cm$^2$, $\beta_\Sigma = 0$,
$t_{\rm disc} = 10^5$ yr), while the resonance
parameters are appropriate for the 2:1 ($\alpha = (1/2)^{2/3}$, $f_1 = -1.190$, and $f_2 = 1.688$). 
At $P_1 < 15$ days (lower panel), 
the observed peak shifts to larger $\Delta$ than is
predicted by our model, and the trough is wider, presumably 
reflecting post-disc tidal dissipation.
}
\label{fig:mc_q1}
\end{figure}

We have focussed so far on using planet-disc
interactions, and not stellar tidal effects, to reproduce
the $\Delta$-distribution near the 3:2 resonance. The 3:2 
data exhibit the strongest peak-trough
asymmetry. These data also appear, as judged by
their relative insensitivity to orbital period
(Figures \ref{fig:delta_psplit} and \ref{fig:cdf}),
the least impacted by stellar tides. 
A non-trivial test of our model
is to see whether it can simultaneously reproduce
the 2:1 $\Delta$-distribution, at least at the longest
periods where potential complications from tides
are minimal. Our model appears to pass this test: \autoref{fig:mc_q1} demonstrates that our best-fit disc model for the 3:2 reproduces the 2:1 
reasonably well at inner planet
periods $P_1 > 15$ days (top panel). 
We emphasize that we have not tuned any of our
disc parameters to fit for the 2:1; we have merely
taken the same background disc
that we fitted for the 3:2 
($\Sigo \sim 20$ g/cm$^2$, $t_{\rm disc} = 10^5$ yr)
and asked whether it reproduces
the 2:1. It does.

We have looked to the parent disc to change planet semi-major axes and
eccentricities. While
\citet[][ML]{millholland_laughlin_2019} also looked to the disc to
drive semi-major axis changes, they appealed to tidal friction, specifically the heat generated by obliquity
tides raised on planets by their host stars, for an additional energy sink to drive resonant repulsion.
All of our results indicate that, at long enough
periods,
the disc suffices as
a source of dissipation.  Eccentricity damping by the disc is neglected
by ML but is part and parcel of the disc-planet
torque.\footnote{Eccentricity damping is effected by first-order
  co-orbital Lindblad resonances, and semi-major axis changes
  (migration) by principal Lindblad resonances (e.g., \citealt{goldreich_sari_2003}).} In concert
with disc-driven migration, disc eccentricity damping creates the peak-trough features at
$\Delta = \pm \, 0.01$ seen at large periods for both the 3:2 and 2:1 resonances, with no need for extra damping.

The picture complicates, however, at the shortest periods.
At $P_1 < 15$ days (bottom panel in
\autoref{fig:mc_q1}), the disc-only, no-tide model is not a good fit
to the 2:1: the observed 2:1 peak is lower in amplitude and displaced
to larger $\Delta$, and the observed 2:1 trough is wider. 
Future work
needs to resolve these discrepancies. 
Planets in the 2:1 are situated farther apart
than in the 3:2, making the 2:1 more prone to disruption
(say by other planets in the system), \rev{and possibly less
prone to convergent migration (see the end of this
section for why); both effects} would reduce the height of the 2:1 
peak. The larger separation also renders 2:1 systems 
less sensitive to their mutual resonant forcing and more sensitive
to stellar tides.
The shift of the location of the peak toward larger
$\Delta$ with decreasing period (Figures \ref{fig:delta_psplit} and \ref{fig:cdf}) 
seems best explained by
tides. In this context the ML scenario specifically calls out the 2:1 over the 3:2: 
ML argued that the parameter space for spin-orbit resonance capture and obliquity-driven
tidal dissipation is larger for the 2:1 than for the 3:2,
and \cite{millholland_2019} uncovered observational evidence for greater
tidal heating of planets in the 2:1 than the 3:2. The emerging qualitative
picture is that planet-disc interactions establish, over Myrs,
a baseline peak-trough asymmetry at all periods (this paper),
while tides take this baseline at the shortest periods
and modify it over Gyrs (ML). 
Asynchronous tides raised on host stars by planets
might also have a role to play---these
cause planets to migrate inward and divergently,
further shifting the peak to larger $\Delta$ 
(\citealt{elee_chiang_2017}, their fig.~10). 

Whereas our model requires only small, $\sim$10\% changes to orbital
periods that are consistent with the lack of observed planet pile-ups
at short period \citep{ogihara_etal_2018, elee_chiang_2017, dressing_charbonneau_2015,
  fressin_etal_2013}, it is not clear whether disc-driven migration in
the ML scenario is similarly compatible.  In the example evolution
shown in fig.~3 of ML, planet orbital periods change by
$\sim$70--80\%: first to cross a spin-orbit resonance, then to capture
into 3:2 resonance, and finally to capture into spin-orbit resonance
and generate a large permanent obliquity. Adjusting initial conditions
and parameters may reduce the degree of migration needed in the ML
scenario. What should also help is an accounting
for how planetary precession rates change
as the disc dissipates (an effect omitted by ML and by us);
explicitly time-varying precession can lead to spin-orbit
resonance crossings with less 
need for semi-major axis changes (see, e.g., \citealt{ward_1981}).

There are other open questions. 
How does the $\Delta$-distribution change
post-disc, over Gyrs of gravitational
interactions between planets? Diffusion
of systems in $\Delta$ would erode the peak-trough
asymmetry, with the lower survival probability of
systems at $\Delta < 0$ compensating in part \citep{pu_wu_2015}. The free eccentricities (resonant libration amplitudes)
of our modeled planets
in the peak are damped to zero by the disc;
how do we raise them to reproduce resonant
systems with observed
free eccentricities of order 1\%
\citep{lithwick_etal_2012,hadden_lithwick_2014,hadden_lithwick_2017}? Here also post-formation interplanetary interactions should be investigated. 
\rev{Finally, most planet pairs in the peak arrived
there in our model
by convergent migration. Whether migration is
convergent or divergent depends on
how the disc is structured (how its aspect ratio
and surface density change with radius),
and the mass ratio $m_2/m_1$ of the outer planet to the inner.
Mass measurements by \citet{hadden_lithwick_2017}
using transit timing variations indicate that 
$m_2/m_1 > 1$ about as often as $m_2/m_1 < 1$ for
planet pairs near resonance peaks.
What do these mass ratio statistics imply about disc structure,
assuming these pairs migrated convergently?
Pairs with $m_2/m_1 < 1$ can still migrate convergently if the surface
density rises with increasing distance from the host
star, as it does in the wind-driven disc models
of \citet{suzuki_etal_2016}. Planetary orbits
also converge regardless of $m_2/m_1$ if a common
gap is opened between them
\citep{masset_snellgrove_2001, fung_chiang_2017}.
The closer the planets, the more easily planetary Lindblad
torques clear a common gap; thus we would expect more convergent pairs capturing into the 3:2 than into the (more separated) 2:1. Indeed the peak for the 3:2 is stronger than for the 2:1.}

\section*{Acknowledgements}
We thank Konstantin Batygin, Rebekah Dawson, Courtney Dressing, Jean-Baptiste Delisle, Paul Duffell, Dan Fabrycky, Sivan Ginzburg, Dong Lai, Yoram Lithwick, Andy Mayo, Sarah Millholland, Masahiro Ogihara, Hanno Rein, and Yanqin Wu for useful exchanges, and Caroline Terquem for a constructive referee report. We also thank
Oleg Gnedin and Dan Weisz for sharing computing resources. 
This work benefited from NASA's Nexus for Exoplanet System Science (NExSS) research coordination network sponsored by the NASA Science Mission Directorate, and was
supported by NASA grant NNX15AD95G/NEXSS. We relied on the Savio computational cluster resource provided by the Berkeley Research Computing program at the University of California, Berkeley (supported by the UC Berkeley Chancellor, Vice Chancellor for Research, and Chief Information
Officer), and the NASA Exoplanet Archive, which is operated by the California Institute of Technology, under contract with NASA under the Exoplanet Exploration Program. We also made use of the \textsc{matplotlib} \citep{hunter_etal_2007} and \textsc{scipy} Python packages.




\bibliographystyle{mnras}
\bibliography{repulsion/planets_nick} 



\appendix

\section{Host Star Spectral Type} \label{sec:app}

\begin{figure*}
\includegraphics[width=\textwidth]{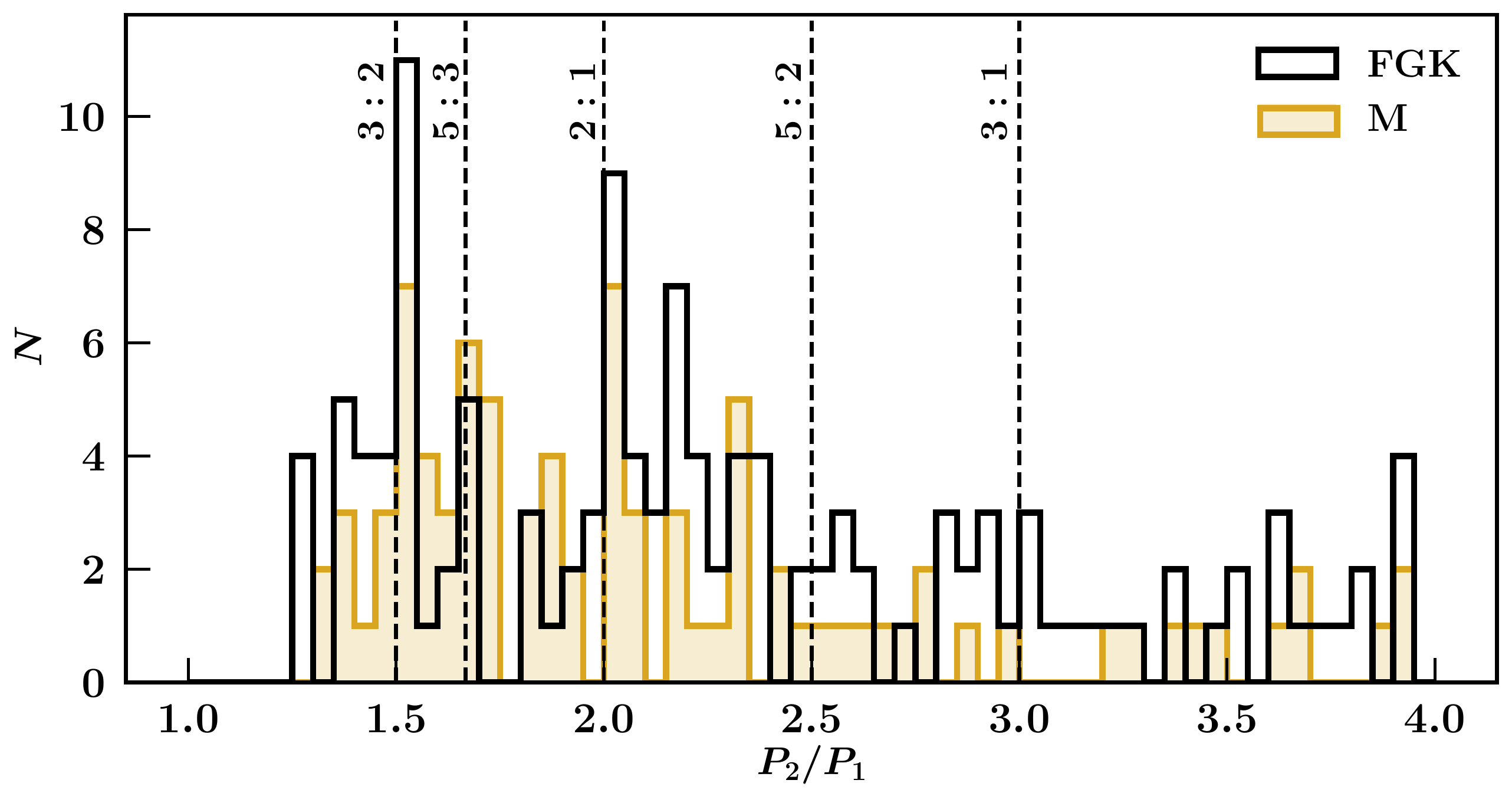}
\caption{Same as \autoref{fig:period_ratio_distributions} but now differentiating between FGK host stars and M stars. The period ratio asymmetries near first-order resonances 
are evident for all spectral types.}
  \label{fig:period_ratios_spectype}
\end{figure*}

In \autoref{fig:period_ratios_spectype} we plot the
distribution of period ratios of pairs of sub-Neptunes,
distinguishing between those with FGK host stars and
those with M host stars.
Compared to \autoref{fig:period_ratio_distributions},
the statistics in \autoref{fig:period_ratios_spectype}
are poorer, 
not only because we are splitting the data but because
we had to discard the many entries in the NASA Exoplanet Archive that do not specify host star spectral type
(\autoref{fig:period_ratio_distributions} plots
all systems regardless of whether they have a spectral
type listed or not). As far as we can tell
from \autoref{fig:period_ratios_spectype},
the peak-trough asymmetries near the 3:2 and 2:1
resonances are common to sub-Neptunes
around both FGK and M stars.

\bsp	
\label{lastpage}
\end{document}